\pdfoutput=1
\documentclass[aps,prd,amsmath,floats,floatfix, twocolumn,
superscriptaddress,nofootinbib,showpacs,longbibliography]{revtex4-1}
\usepackage{xspace}
\usepackage[T1]{fontenc}
\usepackage[utf8]{inputenc}
\DeclareUnicodeCharacter{2009}{\nobreakspace}
\usepackage{lmodern}
\usepackage{times}
\usepackage[varg]{txfonts}
\usepackage[normalem]{ulem}
\usepackage{verbatim}
\usepackage[dvipsnames, usenames]{xcolor}
\definecolor{linkcolor}{rgb}{0.0,0.3,0.5}
\usepackage[hypertexnames=false, unicode, colorlinks=true, linkcolor=linkcolor,
citecolor=linkcolor, filecolor=linkcolor,urlcolor=linkcolor,
pdfusetitle]{hyperref}
\usepackage[all]{hypcap}
\usepackage{graphicx}
\usepackage{xspace}
\usepackage{amssymb}
\usepackage{bm}
\usepackage{microtype}
\usepackage[english]{babel}
\usepackage{orcidlink}
\usepackage{graphicx}
\graphicspath{%
  {Plots}%
}
\usepackage{enumitem}
\DeclareMathAlphabet{\mathpzc}{OT1}{pzc}{m}{it}

\newcommand{\dd}{\text{d}}

\newcommand{\spinvec}{\mathbf{S}\xspace}

\newcommand{\Lvec}{\mathbf{L}\xspace}
\newcommand{\tilt}{\theta^{\spinvec\Lvec}\xspace}
\newcommand{\bilby}{\texttt{Bilby}~\cite{Ashton:2018jfp}\xspace}

\usepackage{titlesec}

\titleclass{\subsubsubsection}{straight}[\subsubsection]
\newcounter{subsubsubsection}[subsubsection]
\renewcommand\thesubsubsubsection{\thesubsubsection.\arabic{subsubsubsection}}
\titleformat{\subsubsubsection}
  {\normalfont\normalsize\bfseries}{\thesubsubsubsection}{1em}{}
\titlespacing*{\subsubsubsection}{0pt}{3.25ex plus 1ex minus .2ex}{1.5ex plus .2ex}

\begin{document}
\title{Eccentricity evolution consistency test to distinguish eccentric gravitational-wave signals from eccentricity mimickers}

\newcommand{\IUCAA}{\affiliation{Inter-University Centre for Astronomy and
    Astrophysics, Post Bag 4, Ganeshkhind, Pune 411007, India}}
\newcommand{\SNU}{\affiliation{Department of Physics and Astronomy,
    Seoul National University, Seoul 08826, Korea}}
\newcommand{\VSM}{\affiliation{Department of Physics, Vivekananda Satavarshiki Mahavidyalaya (affiliated to Vidyasagar University),\\ Manikpara 721513, West Bengal, India}}

\author{Sajad A. Bhat\,\orcidlink{0000-0002-6783-1840}}
\email{sajad.bhat@iucaa.in}
\IUCAA
\author{Avinash Tiwari\,\orcidlink{0000-0001-7197-8899}}
\email{avinash.tiwari@iucaa.in}
\IUCAA
\author{Md Arif Shaikh\,\orcidlink{0000-0003-0826-6164}}
\email{arifshaikh.astro@gmail.com}
\VSM
\author{Shasvath J. Kapadia\,\orcidlink{0000-0001-5318-1253}}
\email{shasvath.kapadia@iucaa.in}
\IUCAA

\date{\today}

\begin{abstract}
Eccentric compact binary coalescences (CBCs) are expected to be
observed in current and future gravitational-wave (GW) detector networks. Such detections are especially valuable as they can provide insights on the environments that nurture CBCs. However, it has been recently pointed out that a number of other physical and beyond-GR
effects, 
could imitate, or be mimicked
by, eccentric CBCs. The standard approach to ascertain that a detected CBC is eccentric is to employ Bayesian model selection, where the eccentric CBC hypothesis is compared against other hypotheses. Such an approach is not only computationally intensive and time-consuming, but could also be misleading if none of the models under consideration represent the true model. In this work, we propose a conceptually
simple but powerful method to directly confirm or reject the eccentric hypothesis, without needing to compare the hypothesis with the plethora of other possible hypotheses. The key idea is that while
spurious non-zero values of eccentricity, at some reference frequency,
could be acquired when a non-eccentric CBC with additional physical/beyond-GR effects is recovered with an eccentric CBC waveform
model, the {\itshape evolution} of eccentricity with frequency will in
general not be mimicked. We accordingly formulate an eccentricity
evolution consistency test (EECT). The method compares the
eccentricity recovered at some low frequency value (e.g, $10$ Hz), evolved to higher frequencies assuming GR, with
eccentricities recovered at those same higher frequencies. Discrepancy between the two eccentricities at any reference frequency would violate EECT and indicate the presence of a mimicker. As a proof of concept, assuming a few eccentric CBC systems, quasi-circular CBCs with additional physics mimicking eccentricity, and an O4-like
three-detector-network configuration, we demonstrate that our proposed method is
indeed able to reject mimickers at $\geq 68\%$ confidence, while ensuring that truly eccentric CBCs satisfy EECT.
\end{abstract}

\maketitle

\section{Introduction}\label{sec:introduction}
Stellar mass compact binary coalescences (CBCs) are one class of
sources that produce detectable gravitational waves
(GWs). Templated searches for these sources in detector data from
LIGO-Virgo-KAGRA's (LVK's) \cite{TheLIGOScientific:2014jea, TheVirgo:2014hva, KAGRA:2020tym} first three observing runs (O1, O2, O3) and the first part of the fourth observing run (O4a) have unearthed $> 200$
confirmed detections~\cite{LIGOScientific:2018mvr,LIGOScientific:2020ibl,KAGRA:2021vkt,LIGOScientific:2025slb,LIGOScientific:2025hdt,LIGOScientific:2025yae}. Most of these pertain to
binary black hole (BBH) mergers~\cite{KAGRA:2021vkt}, although binary
neutron star (BNS)~\cite{TheLIGOScientific:2017qsa, Abbott:2020uma}
and neutron star black hole (NSBH)
mergers~\cite{LIGOScientific:2021qlt} have also been observed.

Identifying the formation channels that produced the observed CBCs is
a high-profile endeavor of considerable
interest~\cite{LIGOScientific:2021psn}. What is becoming increasingly
clear is that a single formation channel almost certainly cannot
explain the provenance of all of the observed CBCs
~\cite{Zevin:2020gbd}. Nevertheless, most of the proposed merger
environments that allow CBCs to merge within Hubble time can be
broadly categorised into two classes: CBCs could either have evolved
in isolation in the galactic field, or have been dynamically assembled
in dense stellar environments~\cite{Zevin:2017evb,Mandel:2018hfr,Taylor:2018iat,Roulet:2018jbe,Baibhav:2020xdf,Mapelli:2021for,Afroz:2024fzp,Stegmann:2025shr,Dorozsmai:2025jlu}.

Ascertaining a CBC's formation channel, on a single event basis, is in
general difficult. This is because there is no incontrovertible
evidence provided by the CBC signal that indicates its formation
channel.~\footnote{A possible exception to this is if the motion of
  the CBC's center of mass imprints itself on the GW waveform (see,
  e.g., \cite{Vijaykumar:2023tjg, Tiwari:2023cpa}), which could be
  extracted to constrain the gravitational potential of the CBC's
  environment.} However, statistical arguments have been made
suggesting that the values of the masses, and the magnitudes and
orientations of the spins of the components of the CBCs, could be used
to acquire hints of their merger environment (see, e.g.,
\cite{Vajpeyi:2021qsw}). Another intrinsic parameter that could
provide additional clues of the CBCs' formation channel is residual
orbital eccentricity at a chosen reference frequency (see, e.g.,
\cite{Gondan:2020svr}).

GWs carry away energy and angular momentum from an inspiralling
compact binary. This circularises the binary \cite{Peters:1963ux}, and
any residual eccentricity of the orbit at the time when the GWs enter
the frequency band of current ground based detectors is expected to be
negligibly small. However, certain formation channels pertaining to
CBCs merging in dense stellar environments could provide detectable
residual eccentricities \cite{Mapelli:2021for}. Dynamical encounters
in such environments could harden the binary sufficiently rapidly so
that the GWs cannot exhaust orbital eccentricity entirely, leaving a
residual amount at band-entry. Furthermore, binary-single encounters
in AGN disks \cite{Samsing:2020tda,tagawa2021eccentric}, and
Kozai-Lidov \cite{Kozai:1962zz, Lidov:1962wjn, Naoz:2016tri,
  Antonini:2017ash} mechanism due to the presence of a third body,
could also boost eccentricity.

Detecting eccentric CBCs promises to shed light on their formation,
environment and evolution. Recently, several gravitational wave events have been claimed to show mild to strong signatures of eccentricity~\cite{Romero-Shaw:2019itr,Romero-Shaw:2021ual,Iglesias:2022xfc,Wu:2020zwr,Planas:2025jny,Morras2025_GW200105_ecc} 
. It is therefore of considerable interest to
confidently determine if a CBC is truly eccentric, or an eccentricity
mimicker. Indeed, several works have shown that certain physical and
beyond-GR effects could masquerade as, or be mimicked by,
eccentricity. For example,
Refs.~\cite{Shaikh:2024wyn,Bhat:2022amc,Saini:2022igm,Saini:2023rto,Narayan:2023vhm}
have shown that tests of general relativity (GR) with quasi-circular
waveforms applied to eccentric CBC signals could result in spurious
violations of GR. On the other hand, an eccentric CBC could mimick one with
finite line-of-sight acceleration \cite{Tiwari:2025aec}, as will be shown later; precession \cite{RomeroShaw2023_precession_eccentricity_degeneracy}; and
one that is microlensed~\cite{mishra-ecc-mclz}.

The prevalent approach to distinguish an eccentric CBC from one that
mimicks it is to use Bayesian model selection. Large scale Bayesian
parameter estimation (PE) is performed on stretches of data containing known CBCs. These PE exercises are carried out under different
hypotheses, and ratios of evidences (Bayes factors) pertaining to
pairs of hypotheses are evaluated to determine the model that best
fits the data. Such an approach
suffers from two problems. The first
is that the model, from the set of models considered, that best fits
the data, will be the true model only if the set contains the true
model. The second is computational time and expense, given the large
number of possible physical and beyond-GR effects that could modulate
a CBC with respect to a quasi-circular one. Evaluating Bayes factors
for all possible pairs becomes increasingly unfeasible with increasing
models in the set.

In this work, we propose an eccentricity evolution consistency test (EECT) that could directly ascertain, or reject, the eccentricity hypothesis, without needing to invoke alternative models and evaluate Bayes factors. The crux of the method rests on the following expectation. While some physical/beyond-GR effects in quasi-circular CBCs, recovered with eccentric models that do not account for these effects, produce spurious non-zero eccentricity at a chosen reference frequency, they will not in general conform to the expected frequency evolution of eccentricity. With this in mind, we first recover eccentricity (and other intrinsic parameters) at a fiducial low frequency (say, $10$ Hz). We then predict the eccentricity at higher frequency values, assuming GR and that the CBC evolves under GW radiation alone . We subsequently recover eccentricities at each of these higher frequency values, and compare the recovered values with the predicted ones. 

As a proof of principle, we test our method on two physical effects and two beyond-GR effects that mimick eccentricity. We assume zero noise with an O4-like noise power spectral density (PSD). We consider non-spinning CBCs, with different sets of component masses that ensure that higher harmonics of GW radiation are suppressed. We find that for all the cases we consider, EECT is violated at $68\%$ or more, thus identifying the mimickers. We also perform null tests to ascertain that the EECT is not violated when a truly eccentric CBC is recovered with an eccentric waveform. 

The rest of the paper is organized as follows. In
  Sec.~\ref{sec:gravitational_wave_basics}, we introduce notations
and conventions, and provide basics of parameter estimation from
GWs. In Sec.~\ref{sec:eccentricity_mimickers}, we discuss potential
eccentricity mimickers, and in
Sec.~\ref{sec:eccentricity_evolution_consistency_test}, we design the
eccentricity evolution consistency test. In Sec.~\ref{sec:results}, we
present our main results, and conclude in Sec.~\ref{sec:conclusion}.
We use geometrized units ($G=c=1$) in the equations 
and present our results in physical units.

\section{Gravitational Waves Basics}\label{sec:gravitational_wave_basics}
\subsection{Notations and conventions}\label{sec:notations_and_conventions}
In GR, the GWs have two polarizations $h_{+}$ (``plus'') and
$h_{\times}$ (``cross''). These polarizations depend on 10 intrinsic
and 4 extrinsic parameters. The intrinsic parameters are the two
detector frame component masses $m_1, m_2$ ($q \equiv m_2/m_1 < 1$), two spin vectors
$\vec{S}_1, \vec{S}_2$ (three components for each compact object),
and two eccentricity parameters (eccentricity $e$ and mean anomaly
$l$)~\footnote{While mean anomaly is the most convenient choice in
many practical cases, other choices for the second eccentricity
parameter~\cite{Clarke:2022fma} like the ``true anomaly'' are also
possible.}. The extrinsic parameters are the luminosity distance
$d_{\rm L}$ of the binary from the detectors, the inclination angle $\iota$
of the binary with respect to the line of sight, the time $t_{\rm c}$ and
phase $\phi_{\rm c}$ of coalescence.

\subsection{Parameter Estimation}{\label{sec:parameter_estimation}}
The GW signal $h$ detected in the ground-based detectors depends on the response
of the detectors to the GW polarizations. The detected signal is a
linear superposition of the polarizations weighted by the antenna
pattern functions of the detectors. The antenna pattern functions
$F_{+}, F_{\times}$ depend on the location of the binary on the sky
(described by the right ascension $\alpha$ and the declination
$\delta$) and the polarization angle $\psi$. Thus, the signal depends
on 17 parameters,~\footnote{This is the minimal number of the parameters required to describe a binary inspiralling as per GR. Beyond-GR theories or external physical effects might require additional parameters.} and can be written as:
\begin{equation}
  \label{eq:signal} h\left(t; \vec{\theta}, \vec{\lambda}\right) =
F_{+}\left(\vec{\lambda}_{\rm F}\right)h_{+}\left(t;\vec{\theta},\vec{\lambda}_{\rm h}\right) +
F_{\times}\left(\vec{\lambda}_{\rm F}\right)h_{\times}\left(t;\vec{\theta},\vec{\lambda}_{\rm F}\right),
\end{equation} where $\vec{\lambda}_{\rm h} = \{m_1, m_2, \vec{S}_1,
\vec{S}_2, e, l\}$ are the 10 intrinsic parameters of the
polarizations, $\vec{\theta} = \{\iota, t_{\rm c}, \phi_{\rm c}, d_{\rm L}\}$ are the 4
extrinsic parameters of the polarizations, and $\vec{\lambda}_{\rm F} =
\{\alpha, \delta, \psi\}$. $\vec{\lambda} = \vec{\lambda}_{\rm h} \cup
\vec{\lambda}_{\rm F}$ and $t$ is the time in the detector frame.

The GW detector data $d(t)$ consists of the signal $h\left(t;\vec{\theta},\vec{\lambda}\right)$ and the detector noise $n(t)$ inside which the signal is buried:
\begin{equation}
    \label{eq:detector_data}
    d(t) = h\left(t;\vec{\theta},\vec{\lambda}\right) + n(t).
\end{equation}
The parameters of the signal $h\left(t;\vec{\theta},\vec{\lambda}\right)$ are estimated from the data $d(t)$ by sampling the 17-dimensional GW posterior $p\left(\vec{\theta},\vec{\lambda}|d\right) \propto p\left(\vec{\theta},\vec{\lambda}\right) p\left(d|\vec{\theta},\vec{\lambda}\right)$, where $p\left(\vec{\theta},\vec{\lambda}\right)$ is the prior distribution on the parameters, and $p\left(d|\vec{\theta},\vec{\lambda}\right)$ is the GW likelihood given by:
\begin{equation}
    \label{eq:likelihood}
    p\left(d|\vec{\theta},\vec{\lambda}\right) \propto \mathrm{exp}\left[ - \frac{\left\langle d - h\left(\vec{\theta},\vec{\lambda}\right), d - h\left(\vec{\theta},\vec{\lambda}\right) \right\rangle}{2}\right].
\end{equation}
The noise-weighted inner product $\langle h_1, h_2\rangle$ between two complex timeseries $h_1$ and $h_2$ is defined as: 
\begin{equation}
    \label{eq:inner_product}
    \langle h_1, h_2\rangle = 4 \mathcal{R} \int_{f_{\mathrm{min}}}^{f_{\mathrm{max}}} \frac{\tilde{h}_1^{*} \left(f;\vec{\theta},\vec{\lambda}\right) \tilde{h}_2\left(f;\vec{\theta},\vec{\lambda}\right)}{S_{\rm n}(f)} \dd f,
\end{equation}
where $f$ is the frequency, $\tilde{h}_1, \tilde{h}_2$ are the Fourier transforms of the timeseries $h_1, h_2$, respectively, $*$ represents the complex conjugate, and $\mathcal{R}$ represents the real part. $S_{\rm n}(f)$ is the noise power spectral density (PSD) of the detector that sets its sensitivity to GW signals. The lower and upper cutoff frequencies, $f_{\mathrm{min}}$ and $f_{\mathrm{max}}$, are determined by the source properties and the sensitivity of the detectors.

The high-dimensional GW posterior is generally sampled using
large-scale Markov Chain Monte Carlo (MCMC) or Nested-sampling methods. In
this work, we utilise the state-of-the-art GW Bayesian parameter inference software
\textsc{Bilby} \cite{Ashton:2018jfp, Smith:2020} along with a dynamical nested sampler \textsc{dynesty}~\cite{Speagle:2019ivv} for GW parameter estimation.

\section{Eccentricity mimickers}\label{sec:eccentricity_mimickers}
\subsection{Features of GWs from a binary in an eccentric orbit}\label{sec:features_of_gws_from_binary_on_eccentric_orbits}
Due to the loss of energy and angular momentum via radiation of GWs,
the orbit of the binary loses eccentricity and becomes circular as 
the binary inspirals. At small eccentricity, the eccentricity follows
a power law decay as a function of the orbital frequency 
$e \propto f_{\mathrm{orb}}^{-19/18}$ \cite{Peters:1963ux}. The expectation therefore is that any residual eccentricity when the GWs enter the detector band of the LVK detector network will be negligible. Thus, the vast majority of works involving GW data analysis typically use  quasi-circular templates.
However, these are not adequate for inferring the parameters of a binary
on orbits with non-negligible residual eccentricity, as expected from certain mergers in dense stellar environments or hierarchical triples. The waveforms describing GWs emitted
by binaries on eccentric orbits have richer morphology compared to the quasi-circular
waveforms.

The GW frequency and the amplitude of a quasi-circular binary increase 
monotonically over time. On the other hand, an eccentric binary produces bursts
of GWs as the individual components traverse the pericenters of their orbits about the mutual center of mass. Consequently,
the waveform frequency and amplitude in the time domain exhibit
modulations over the orbital timescale with peaks at the pericenters
and troughs at the apocenters (location of the farthest
separation). These are reflected in the frequency domain waveform as a  
frequency-dependent modulation.

In recent times, it has become evident that other physical/beyond-GR
effects may also introduce frequency-dependent modulations in the
gravitational waveforms. Therefore, these effects may be degenerate
with the eccentric effect for certain systems, which may lead to
falsely identifying these systems as eccentric ones. Below, we
discuss four such eccentricity mimickers.

\subsection{Microlensed GWs}\label{sec:microlensed}
Microlensing of GWs occurs when compact
objects—such as stars or stellar-mass black holes with Schwarzschild
radius $R_{\rm Sch}$ comparable to the wavelength $\lambda_{\rm GW}$ of
GWs—intervene between the GW source and the detector, producing images that are unresolved in time, exhibiting wave-optics effects such as interference and diffraction. These result in frequency-dependent modulations of the
waveform. In the frequency domain, a standard unlensed \texttt{TaylorF2}
waveform of a quasi-circular binary inspiral, given by \cite{Buonanno:2009zt} : 
\begin{equation}
\tilde{h}(f) = A(f)\, e^{-i\Psi(f)},
\end{equation}  
is modified by a complex amplification factor \( F(\omega, y) \), resulting in the lensed waveform:  
\begin{equation}
  \label{eq:lensed_h}
\tilde{h}_{\rm lensed}(f) = F(\omega, y)\, \tilde{h}(f). 
\end{equation}  
Here, \( \omega = 8\pi  M_{\rm L} (1+z_{\rm L}) f \) is the dimensionless
frequency, where \( M_{\rm L} \) is the lens mass and \( z_{\rm L} \) is the lens
redshift. The parameter \( y =
\beta/\theta_{\rm E} \)\footnote{$y$ may also be interpreted as the impact parameter.} denotes the scaled 
source–lens angular separation, with Einstein angle:   

\begin{equation}
  \label{eq:thetaE}
  \theta_{\rm E} = \sqrt{ 4  M_{\rm L} \frac{D_{\rm LS}}{D_{\rm L} D_{\rm S}} },
\end{equation}
defined in terms of angular-diameter distances \( D_{\rm L} \), \( D_{\rm S} \), and \( D_{\rm LS} \). For a point-mass lens, the amplification factor \( F(\omega, y) \) is given by \citep{Takahashi:2003ay, Dai:2017huk, Yeung:2021chy,Mishra:2021xzz}:  
\begin{align}
F(\omega, y) & = \exp\left[ \frac{\pi \omega}{4} + \frac{i\omega}{2}\left(\ln\left(\frac{\omega}{2}\right) - 2\phi_m(y)\right) \right] \nonumber \\
& \times \Gamma\left(1 - \frac{i\omega}{2}\right)\, {}_1F_1\left(\frac{i\omega}{2}, 1; \frac{i\omega}{2} y^2 \right),
\end{align}
where:
\[
\phi_m(y) = \frac{(x_m-y)^2}{2}- \ln(x_m)
\]
is the dimensionless Fermat potential difference with $x_m=(y+\sqrt{y^2+4})/2$ and ${}_1F_1$ is the confluent hypergeometric function. The real modulus \(
|F(\omega, y)| \) introduces a frequency-dependent amplification,
while \( \arg F(\omega, y) \) contributes a phase shift. The resulting waveform modulations have been shown to exhibit degeneracies with eccentricity driven modulations \cite{mishra-ecc-mclz}. Indeed, as will also be seen in this work, microlensed GWs recovered with eccentric templates result in spurious non-zero estimates of eccentricity at a given reference frequency. 

\subsection{Line Of Sight Acceleration (LOSA)}\label{sec:losa}
Line-of-sight acceleration (LOSA) refers to the constant acceleration \(a\) of a compact binary’s center of mass along the observer’s line of sight.
It induces a secular time-dependent Doppler shift in the observed GW signal, producing an additional frequency-dependent phase term. In frequency-domain models like the \texttt{TaylorF2} waveform, LOSA leads to a leading-order correction that appears at relative \(-4\) post-Newtonian (PN) order \footnote{Note that $(n/2)$PN order term in the GW phasing refers to the term proportional to $v^n$ relative to the dominant Newtonian-term scaling as $v^{-5}$.},
scaling as \(v^{-8}\) relative to the dominant quadrupole
term~\cite{Bonvin:2016qxr,Inayoshi:2017hgw,Vijaykumar:2023tjg}, where $v= (\pi M f)^{1/3}$ is the orbital velocity expressed in terms of observed GW frequency $f$ and total mass $M$ of the binary. The
corrected GW phase becomes: 
\begin{equation}
\Psi(f) = \Psi_{\mathrm{TF2}}(f) + \delta\Psi_{\mathrm{LOSA}}(f),
\end{equation}
with the LOSA contribution given by:
\begin{equation}
\delta\Psi_{\mathrm{LOSA}}(f) =  \frac{25}{65536} \left(\frac{M}{\eta^2}\right)\,a\, v^{-13},
\end{equation}
where \(\eta = m_1 m_2 / M^2\) is the symmetric mass ratio, and \(M = m_1 + m_2\) is the detector-frame total mass of the binary. Note that, when presenting LOSA in physical units, it is typically expressed as $a/c$ in units of ${\rm s}^{-1}$ \cite{Tiwari:2025aec, Vijaykumar:2023tjg}.
Since the correction term scales as \(v^{-13}\), it grows rapidly at low frequencies and can accumulate significantly in long-duration observations, particularly for low-mass binaries. 

The enhancement of LOSA-induced phase corrections at low frequencies is similar to orbital eccentricity, where correction to the waveform phase starts at ${-19/6 \, \rm PN}$ ~\cite{Moore:2016qxz}, and raises the possibility of parameter degeneracy between LOSA and eccentricity. Accurate parameter inference thus requires studying this degeneracy in detail to avoid wrongly attributing LOSA-induced phase shifts to eccentric orbits, or vice versa, especially in hierarchical triple systems or dense clusters where both effects may coexist~\cite{Randall:2018nud,Tiwari:2023cpa}.

\subsection{Massive graviton}\label{sec:mg}

The \textit{massive graviton effect} arises in modified gravity theories where the graviton has a nonzero rest mass. This causes GWs to propagate at a frequency-dependent speed, introducing dispersion into the signal as it travels from source to detector. The modified dispersion relation can be written as $E^2 = p^2 + A_0$ as opposed to the classical relativistic $E^2 = p^2$. Here, $A_0 = m_{\rm g}^2$, $m_{\rm g}$ is the mass of the graviton, and $E$ and $p$ are the energy and momentum of the GWs, respectively. In frequency domain waveform models like \texttt{TaylorF2}, this dispersion leads to a correction in the GW phase that enters at $1$PN order relative to the leading Newtonian term. The correction to the phase is given by~\cite{PhysRevD.100.104036, Will:1997bb}: 
\begin{equation}
    \label{eq:phase_corr_mg}
    \Delta \Psi_{1, \rm MG} = - {\rm sign}(A_0) \frac{\pi d_{\rm L}^2 (1 + z)}{d_0} \frac{1}{f} \frac{1}{\lambda_0^2}
\end{equation}
where $d_{\rm L}$ is the luminosity distance, $\lambda_0 = h/\sqrt{\vert A_0 \vert} = h/m_{\rm g}$ is the Yukawa screening length, $h$ is Planck's constant, $z$ is the cosmological redshift of the binary, and $d_0$ is given by Eq. 5 of \cite{PhysRevD.100.104036}.

 In Will’s original derivation~\cite{Will:1997bb}, a matter-only cosmology was assumed, but for modern analyses this is generalized to $\Lambda$CDM~\cite{Yunes:2011ws}. Since the massive graviton phase correction is stronger at lower frequencies, it mimics the effect of \textit{orbital eccentricity}, which also modifies the early inspiral phasing~\cite{Yunes:2011ws}. This shared influence on the low-frequency regime introduces a potential degeneracy between eccentricity and massive graviton effects, which must be carefully disentangled in waveform modeling and parameter estimation to enable unbiased parameter recovery and robust tests of GR.

\subsection{Dipole radiation}\label{sec:dipole}
The conservation of the matter stress-energy tensor in GR allows only quadrupole or higher-order multipole moments to contribute to the GW emission. However, in a number of beyond-GR theories, monopole and dipole emissions are also allowed due to the matter stress-energy tensor not being conserved \cite{Will:2014kxa}. Assuming that $B$ is a theory-dependent parameter that can regulate the strength of the dipole term, the phase correction due to the dipole radiation can be written as \cite{PhysRevLett.116.241104, PhysRevD.86.022004}: 
\begin{equation}
    \label{eq:phase_corr_dr}
    \delta \Psi_{\rm DR}(f) = - \frac{3}{224\, \eta } B\, v^{-7}
\end{equation}
where $v$ is the post-Newtonian parameter defined in previous sections and $\eta$ is the symmetric mass ratio.
 Dipole radiation, predicted by many alternative theories of gravity (such as scalar-tensor models), introduces an additional channel of energy loss that accelerates the inspiral of compact binaries. Similarly, eccentricity hastens orbital energy loss, potentially leading to degeneracies in the modulations (to quasi-circular waveforms) driven by dipole radiation and eccentricity. We indeed find that finite eccentricities are estimated when eccentric waveform models are used to recover GW signals with signatures of dipole radiation. 
 
\section{Eccentricity Evolution Consistency Test (EECT)}\label{sec:eccentricity_evolution_consistency_test}
Knowing the fact that eccentricity-induced frequency modulations in
the otherwise quasi-circular waveform can potentially mimic or be
mimicked by several other physical and beyond-GR effects in the same
waveform, there is a danger of claiming observation of a spuriously
eccentric gravitational wave event~\cite{Romero-Shaw:2022fbf,
  mishra-ecc-mclz,Bhat:2022amc}. To address this problem without requiring to invoke multiple different waveform models, we propose an eccentricity evolution consistency test (EECT) that assesses the consistency of the GW signal with GR-consistent frequency evolution of eccentricity. 

As a proof of concept, we consider inspiral-only non-spinning CBCs. For eccentric orbits, we use the {\tt TaylorF2Ecc} waveform~\cite{Moore:2016qxz}. Using Bayesian parameter inference with {\tt TaylorF2Ecc} as the model, we construct a posterior on the eccentricity $e_0$ -- at reference frequency $f_0$ -- of a GW signal buried in noise. We evolve $e_0$ (and its corresponding posterior) to another fiducial reference frequency to acquire $e_{\rm GR}$ (and its corresponding posterior), using the ${\rm 3PN}$-accurate GR-predicted
frequency evolution given by ~\cite{Moore:2016qxz}:
\begin{equation}\label{ecc-evol}
e(f) = e_0 \left( \frac{f_0}{f} \right)^{19/18} \frac{{\mathcal E}(\xi_{\phi})}{{\mathcal E}(\xi_{\phi,0})},
\end{equation}
where:
\begin{equation}
\label{eq:E}
{\mathcal E}(\xi_{\phi}) = \left[1+\left(-\frac{2833}{2016}+\frac{197 \eta }{72}\right) \xi_{\phi} ^{2/3}  + \cdots + \mathcal{O}(\xi_\phi^{7/3}) \right] \,, \nonumber
\end{equation}  $\xi_{\phi}= \pi M f$ and $\xi_{\phi,0}= \pi M
f_0$. Here $M$, $\eta$ are, respectively, the total mass and symmetric
mass ratio of the compact binary. We then once again infer the eccentricity ($e_{\rm obs}$) at this new reference frequency. 

To evaluate the difference between the predicted and observed eccentricities at the new reference frequency, we propose a deviation parameter $\delta_e (f)$ defined as: 
\begin{equation}
\label{consist-1}
\delta_e (f) = 2 \frac{e_{\rm GR}(f)-e_{\rm obs}(f)}{e_{\rm GR}(f)+e_{\rm obs}(f)}\,,
\end{equation}
where $e_{\rm GR}(f)$, $e_{\rm obs}(f)$ are, respectively, the GR-predicted and observed eccentricity of the binary corresponding to GW frequency $f$.
This parameter quantifies the fractional difference between the two estimates of eccentricity. If the GW signal is truly eccentric, the posterior on the deviation parameter, constructed from the posteriors on $e_{\rm GR}$ and $e_{\rm obs}$, should be consistent with zero within a credible interval, which we set to $68\%$. If not, EECT is violated at $\geq 68\%$ confidence.

We set the initial reference frequency to $f_0 = 10$ Hz, and consider multiple new reference frequencies $f_{i}=
15,\,20,\,..\,,\,40\, {\rm Hz}$ at which to evaluate $\delta_e$. Violation of EECT at any of these frequencies will be a tell-tale sign of an inconsistency between the GW signal and the eccentric waveform model used to estimate the parameters of the source \footnote{Note that, by design, EECT is always satisfied at $f_i = f_0 = 10$ Hz}. This inconsistency could stem from the fact that the GW signal is in fact not eccentric, but contains modulations with respect to quasi-circular waveforms due to additional physical or beyond-GR effects. Indeed, deviations from zero of $\delta_e$ due to such effects typically exhibit a systematic degradation of consistency with increasing reference frequency (see Sec.~\ref{sec:results}).

A few important points should be noted.
\begin{enumerate}[label=\alph*)]
    \item {\it Prior Reweighting} -- Evaluating the posterior on $\delta_e$ self-consistently, from the posteriors on $e_{\rm GR}$ and $e_{\rm obs}$, requires priors pertaining to these posteriors to be equivalent. To that end, we reweight the evolved posterior on $e_{\rm GR}$ by the ratio of the prior on $e_{\rm obs}$ to the evolved prior on $e_{\rm GR}$. This is similar to what is done in the Inspiral Merger Ringdown Consistency Test (IMRCT) \cite{Ghosh:2016qgn,LIGOScientific:2021sio} where the evolved posterior on the final mass and spin of the merged BBH predicted from the inspiral is reweighted to the prior used for the measured posterior on the final mass and spin. 
    
    \item {\it Maximum Reference Frequency} -- It is crucial to determine the maximum reference frequency at which the distribution on $\delta_e$ is evaluated. This is because small eccentricity values ($e \lesssim 0.01$ in O4) tend to be recovered with significant railing of the posterior against the lower limit of the prior at $e=0$. This could result in spurious violations of EECT for eccentric signals, and exaggerate EECT violation for mimickers. We therefore restrict ourselves exclusively to those reference frequencies at which $e = 0$ is excluded at $\geq 68\%$ confidence. 

    \item {\it Minimum and Reference Frequency} -- It is straightforward to see that, simply changing the reference frequency at which $e$ is measured, without changing the frequency range over which the waveform is fitted to the data (or, equivalently, over which the likelihood is evaluated), will result in EECT being {\it always} satisfied. This is a direct consequence of the eccentric waveform model used for recovery (in our work, {\tt TaylorF2Ecc}) being internally consistent with GR. We therefore also change the range over which the likelihood is evaluated by setting the minimum frequency to the reference frequency ($f_{\rm min} = f_i$).

\end{enumerate}

\section{Results}\label{sec:results}
We present results of null tests -- where eccentric signals are shown to statisfy EECT, as well as EECT applied to mimickers. We consider zero-noise and a O4-like scenario involving two LIGO detectors and one Virgo detector. The noise PSDs are taken from \href{https://dcc.ligo.org/LIGO-T2000012-v2/public}{https://dcc.ligo.org/LIGO-T2000012-v2/public}\footnote{Specific files used as noise PSD are \texttt{aligo\_O4high.txt}, \texttt{avirgo\_O4high\_NEW.txt}, and \texttt{kagra\_80Mpc.txt}.}.

We use \texttt{TaylorF2Ecc}\footnote{Set to be 3PN-accurate in this work.} for the eccentric waveform model, and \texttt{TaylorF2} for the quasi-circular model -- to which other physical or beyond-GR effects are tacked on as either a complex frequency-dependent amplification factor, or corrections to the PN expansion of the \texttt{TaylorF2} phase. We assume the binaries to be non-spinning throughout, and consider component masses that ensure that the higher-modes content in the signals are negligible \footnote{Given that this work is a proof-of-concept of EECT, the simplified scenarios considered here were chosen to reduce computational cost.}. 

\subsection{Null-test}\label{sec:null-test}

\begin{table}
\begin{ruledtabular}
\begin{tabular}{@{}c@{}c@{}c@{}}
  Parameter & Injection & Prior \\\hline
  $\mathcal{M}_c (M_\odot)$ & ${\{3.045,\, 14.292,\, 28.095,\, 49.99\}}$ & $\mathcal{U}(1, 300)$\\
  $q$ & $\{0.5,\, 0.805,\, 0.83\,\, 0.916\} $ & $\mathcal{U}(0.1, 1)$\\
  $\chi_{1}$ & 0 & 0 \\
  $\chi_{2}$ & 0 & 0 \\
  $\tilt_1$ (rad) & 0 & 0 \\
  $\tilt_2$ (rad) & 0 & 0 \\
  $d_{\rm L}$(MPc) & $\{40,\, 80,\, 100,\, 200,\, 400,\,1000,\,1800\} $ & \text{PL}$(2, 10, 5000)$\\
  $\iota$ & 2.009 & $\sin(0, \pi)$\\
  $\alpha$(rad) & 3.401 & $\mathcal{U}(0, 2\pi)$\\
  $\delta$ & -0.335 & $\cos(-\pi/2, \pi/2)$\\
  $\psi$(rad) & 0.054 & $\mathcal{U}(0, \pi)$\\
  $t_{\rm c}$(s) & 1187008882.43 & $\mathcal{U}(t_0-0.13, t_0+0.07)$\\
  $\phi_{\rm c}$(rad) & 6.052 & $\mathcal{U}(0, 2\pi)$\\
  $e$ & $\{0.16,\, 0.20,\, 0.27\} $ & $\mathcal{U}(10^{-6}, 0.9)$\\
\end{tabular}
\end{ruledtabular}
\caption{\label{tab:null-prior} Injection values and priors used for parameter estimation with
\bilby for both the eccentric and eccentricity mimicker injections. $\mathcal{U}(a, b)$ stands for a uniform prior in the range $(a,
b)$ and PL($\alpha,a,b$) stands for Power-Law of index $\alpha$ on the range ($a,b$). $\mathcal{\sin}(0, \pi)$ and $\mathcal{\cos}(-\pi/2, \pi/2)$ represent sine and cosine priors over their respective ranges. $t_0 \equiv 1187008882.43$ represents the geocent time, and values like 0 indicate Dirac delta priors. The injection value is mentioned for only those parameters which remain fixed throughout, and injections for variable parameters are mentioned in the text at the relevant places.}
\end{table}

\begin{figure*}[th]
  \centering
  \includegraphics[width=\textwidth]{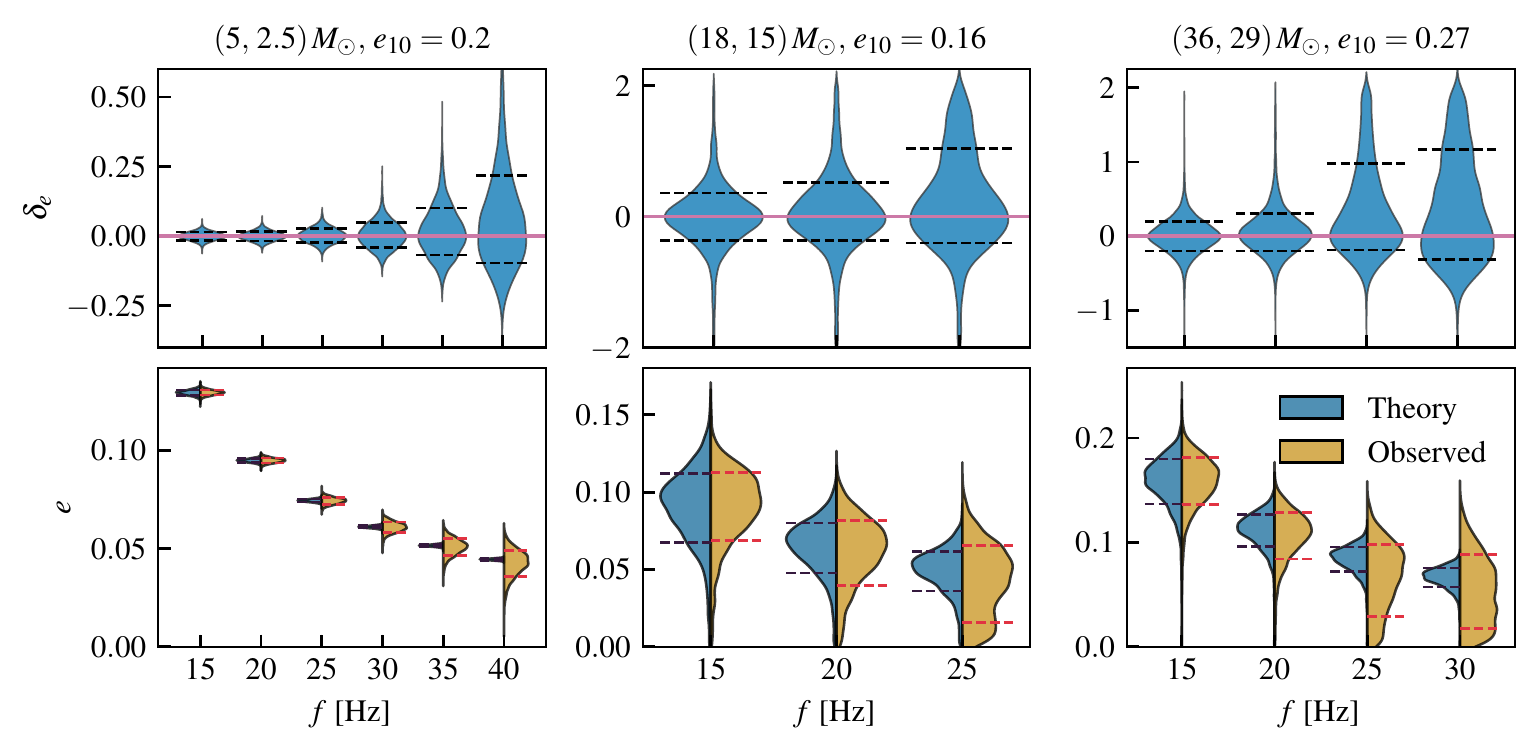}
  \label{fig: null_test_diff_post}
  \caption{\textit{Top}: Violins representing eccentricity deviation $\delta_{e}$ are plotted as a function of GW frequency for null-test wherein we inject eccentric GW signal using {\tt TaylorF2Ecc}
    waveform model in zero-noise and recover the binary parameters
    of the GW signal using the same waveform model. \textit{Bottom}: Individual GR-predicted (left-half) and observed (right-half) eccentricity (i.e, $e_{\rm GR}$ and $e_{\rm obs}$) half-violins are plotted as a function of GW frequency. \textit{Left}, \textit{middle} and \textit{right} panels in each row correspond to binaries with parameters ($m_1=5\,M_{\odot},\,m_2= 2.5\, {M_{\odot}}$, $d_{\rm L}= 100$ Mpc, $e_{10}=0.20$), ($m_1=18\,M_{\odot},\,m_2= 15\, {M_{\odot}}$, $d_{\rm L}= 400$ Mpc, $e_{10}=0.16$) and ($m_1=36\,M_{\odot},\,m_2= 29\, {M_{\odot}}$, $d_{\rm L}= 1000$ Mpc, $e_{10}=0.27$), respectively.}
\end{figure*}

We consider three representative binary
systems, a lighter system with total mass $M = 7.5\, {M_{\odot}}$
($q = 0.5$), a medium mass system with $M = 33\,
{M_{\odot}}$ ($q=0.83$) and a third heavier {\tt
  GW150914}-like  system with $M = 65\,
{M_{\odot}}$ ($q = 0.8055$). These are placed at representative luminosity distances of
$100\, {\rm Mpc}$, $400\, {\rm
  Mpc}$ and $1000\, {\rm Mpc}$, and 
are constructed to have eccentricities of $e_{10}=0.20, 0.16, 0.27$, respectively, at a reference frequency
$f_i = f_0 = 10\, {\rm Hz}$. We infer the eccentricities $e_{\rm
  obs}$ at different reference (and minimum) frequencies ($f_i= 10,\,
15,\,20,\, ..,\, 40\, {\rm Hz}$) by sampling their GW posteriors \footnote{Across all analyses in this {\it paper}, we keep the maximum frequency fixed to $2048 \, \rm Hz$.}. The corresponding priors are tabulated in Table~\ref{tab:null-prior}. The SNRs of the recovered signals for each of the systems with total masses $7.5\,{M_{\odot}}$, $33\,{M_{\odot}}$, $65\,{M_{\odot}}$ are given as $(36.2,\,32.7)$ corresponding to $f_{\rm min}=(10,\,40\, {\rm Hz})$, $(32.8,\,32.4)$ corresponding to $f_{\rm min}=(10,\,25\, {\rm Hz})$, $(23.02,\,22.20)$ corresponding to $f_{\rm min}=(10,\,30\, {\rm Hz})$, respectively. 

We evaluate the GR-predicted eccentricities from the inferred posterior on $e_{10}$, and construct the corresponding evolved posteriors on $e_{\rm GR}$ at other
reference frequencies. We do so using the GR-consistent 3PN-accurate frequency-evolution of
eccentricity as given in Eq.~\eqref{ecc-evol}. 

In Figure~(\ref{fig: null_test_diff_post}), we present the results of
EECT for the null-tests on the three systems mentioned above. The \textit{top} row
shows the posterior on the  eccentricity deviation parameter $\delta_e$ (cf. Eq.~\ref{consist-1}), constructed from the posteriors on $e_{\rm GR}$ and $e_{\rm obs}$, evaluated at reference frequencies $f_i$. 
The \textit{bottom} row displays the individual posteriors on $e_{\rm GR}$ and $e_{\rm obs}$. Different columns correspond to each of the three different CBC systems.
The $\delta_{e}$ values are consistent with
zero at the $68\,\%$ credibility level for all $f_i$ considered. 

It is interesting to note that the widths of the $e_{\rm GR}$ posteriors decrease with increasing frequency. This is a direct consequence of larger eccentricity values decaying more rapidly with frequency than smaller ones \cite{Peters:1963ux}. However, the ratio of the widths of the posteriors (e.g., the standard deviation) to the median value, does not evolve appreciably with frequency.

\subsection{EECT applied to eccentricity mimicker injections}\label{sec:mimicker-test}

We  consider potential eccentricity mimickers which broadly fall
into two categories, i.e., physical effects within GR and beyond-GR
effects. In Sec.~\ref{sec:ecc-mimicker-mclz} and
  Sec.~\ref{sec:ecc-mimicker-losa}, we discuss EECT on eccentricity
  mimickers due to physical effects---microlensing of GWs and LOSA of
  the binary's center of mass, respectively. In
  Sec.~\ref{sec:ecc-mimicker-MG} and \ref{sec:ecc-mimicker-DR}, we
  discuss EECT on eccentricity mimickers due to beyond-GR
  effects---massive graviton and dipole radiation, respectively.

\subsubsection{EECT on microlensed signal}\label{sec:ecc-mimicker-mclz}

\begin{table*}
\begin{ruledtabular}
\begin{tabular}{@{}c@{}c@{}c@{}c@{}c@{}}
  Parameter & Prior (Microlensing) & Prior (LOSA) & Prior (Massive graviton) & Prior (Dipole radiation) \\\hline
  $\mathcal{M}_c (M_\odot)$ & $\mathcal{U}(20, 40)$ & $\mathcal{U}(2.5, 3.5)$ & $\mathcal{U}(20, 100)$ & $\mathcal{U}(1, 300)$ \\
  $q$ & $\mathcal{U}(0.1, 1)$ & $\mathcal{U}(0.05, 1)$ & $\mathcal{U}(0.1, 1)$ & $\mathcal{U}(0.1, 1)$ \\
  $d_{\rm L}$(MPc) & PL$(2, 1, 3000)$ & PL$(2, 10, 1000)$ & PL$(2, 500, 5000)$ & PL$(2, 1, 3000)$ \\
  $e$ & $\mathcal{U}(10^{-6}, 0.5)$ & $\mathcal{U}(0, 0.4)$ & $\mathcal{U}(10^{-6}, 0.9)$ & $\mathcal{U}(10^{-6}, 0.9)$ \\
\end{tabular}
\end{ruledtabular}
\caption{\label{tab:mimicker-prior} Priors used for parameter estimation with
\bilby for eccentricity mimicker injections. $\mathcal{U}(a, b)$ stands for a uniform prior in the range $(a,
b)$ and PL($2,a,b$) stands for Power-Law of index $2$ in the range ($a,b$). Note that the priors used for parameters not mentioned here are the same as for eccentric injection in Table~\ref{tab:null-prior} except for the prior on $t_{\rm c}$ in case of LOSA injection which is a Dirac delta centered at $t_0 \equiv 1187008882.43$.}
\end{table*}

\begin{figure*}[ht]
    \centering
    \includegraphics[width=\textwidth]{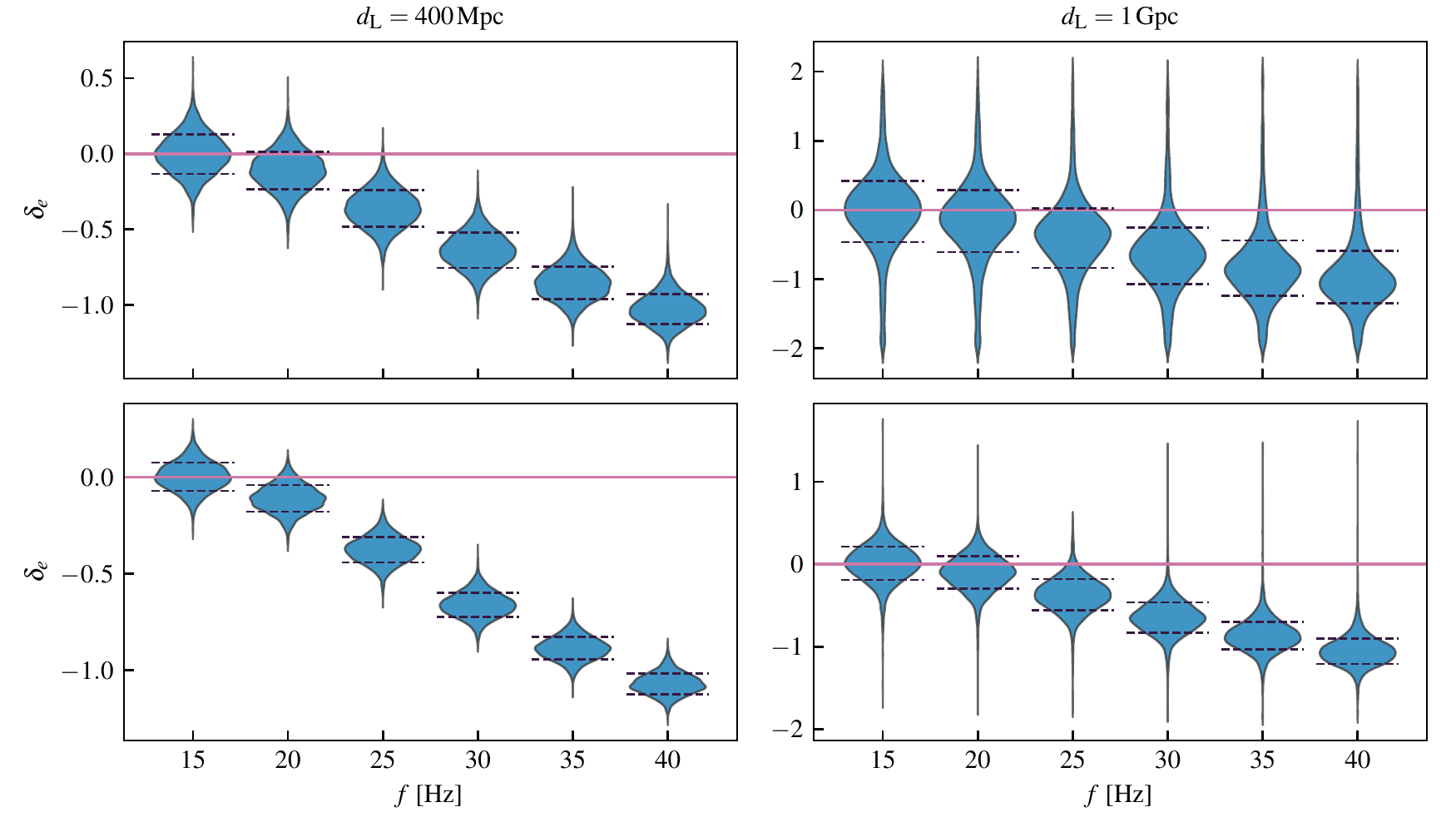}
    \label{fig: Mcl_diff_post}
    \caption{Violins representing eccentricity deviation $\delta_{e}$ are plotted as a function of GW frequency for the case of a non-spinning quasi-circular microlensed CBC injected in zero-noise. The microlensing injection parameters, i.e., redshifted lens-mass $M_{\rm lz}$ and dimensionless impact parameter $y$ are chosen to be $300\, {M_{\odot}}$ and $0.1$, respectively. Here \textit{top} and \textit{bottom} rows correspond to binary of total mass $33\, {M_{\odot}}$ (with mass ratio $q= 0.83$) and {\tt GW150914-like} system of total mass $65\, {M_{\odot}}$ (with mass ratio $q= 0.805$), respectively. The \textit{left} and \textit{right} columns correspond to luminosity distances of $400\, {\rm Mpc}$ and $1\, {\rm Gpc}$, respectively. In all cases, EECT is found to be violated at $\geq 68\%$ beyond a reference frequency.}
\end{figure*}

\begin{figure*}
    \centering
    \includegraphics[width=\textwidth]{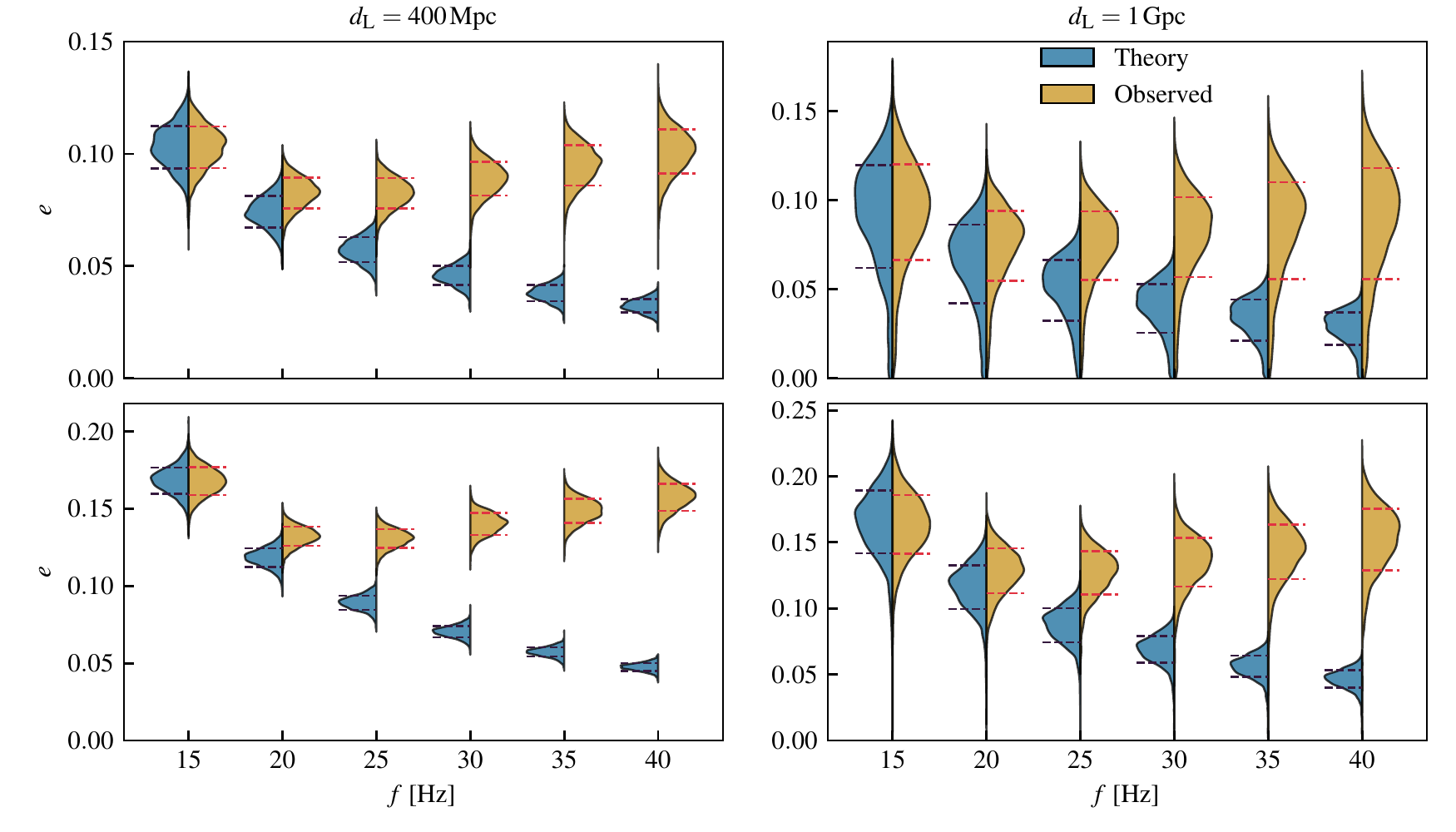}
    \label{fig: Mcl_dv}
    \caption{Same as Figure~\ref{fig: Mcl_diff_post} except that here individual GR-predicted ($e_{\rm GR}$, left-half) and observed ($e_{\rm obs}$, right-half) eccentricity half-violins are plotted as a function of GW frequency. The increasing divergence between the Theory and Observed posteriors with increasing frequency is a tell-tale sign that EECT is being violated in the presence of a mimicker.}
\end{figure*}
 
For microlensing injection, we use the {\tt TaylorF2} waveform
corrected for microlensing amplification as discussed in
Sec.~\ref{sec:microlensed}. The microlensing parameters, i.e., the
redshifted lens-mass ${M_{\rm lz}}$ and the impact parameter $y$ are fixed to be
$300\,{M_{\odot}}$ and $0.1$, respectively. We
consider two representative compact binary systems, a lighter system
with a total mass of $33\, {M_{\odot}}$ (with mass ratio $q=
0.83$) and the other is heavier {\tt GW150914-like} system with total mass $65\, M_{\odot}$ and mass ratio $0.8055$. Each of
the binary systems is considered at two representative luminosity
distances of ${ d_{\rm L}}= 400,\, 1000\, {\rm MPc}$. We recover the parameters of the injected microlensed quasi-circular
signal using unlensed {\tt TayloF2Ecc} 
waveform, thereby acquiring posteriors on $e_{\rm obs}$ at reference frequencies $f_i$. As with the null tests, we also evaluate the posteriors on $e_{\rm GR}$ evolved from the $e_{\rm obs}$ posterior at $f_i = f_0 = 10$ Hz, as well as the posteriors on the deviation parameters $\delta_e$. Priors on the binary parameters, pertaining to the $e_{\rm obs}$ posteriors, are tabulated in Table~\ref{tab:mimicker-prior}. The SNRs of the recovered signals, for injections with $d_{\rm L} = 400 \, {\rm Mpc}$, corresponding to $f_{\rm min}=(10\,{\rm Hz},\,40\, {\rm Hz})$ are given as $(95.89,\,93.41)$ and  $(168.47,\,164.05)$ for the systems with total mass  $33\,{M_{\odot}}$ and $65\,{M_{\odot}}$, respectively, while the same for $d_{\rm L} = 1 \, {\rm Gpc}$ injections are given as $(38.22,\,37.22)$ and $(67.32,\,65)$.

The variation of the posteriors on $\delta_e$ with reference frequencies are shown in Figure~(\ref{fig: Mcl_diff_post}) and
the corresponding individual frequency-evolutions of $e_{\rm GR}$ and
$e_{\rm obs}$ are shown in Figure~(\ref{fig: Mcl_dv}). We observe
that for the lighter system at a luminosity distance of $400\, {\rm Mpc}$
($1\, {\rm Gpc}$) there is a clear violation of the EECT at $68\, \%$
credibility for reference frequencies $f_{i} \ge 25\, {\rm Hz}$
($f_{i} \ge 30\, {\rm Hz}$). While for the heavier {\tt
  GW150914-like} system at a luminosity distance of $400\, {\rm Mpc}$
($1\, {\rm Gpc}$), EECT is violated  at $68\, \%$ credibility for
reference frequencies $f_{i} \ge 20\, {\rm Hz}$ ($f_{i}
\ge 25\, {\rm Hz}$). These results demonstrate how EECT is
successfully identifying eccentricity mimickers, to wit: CBCs microlensed by point mass lenses. 

\subsubsection{EECT on signal with LOSA effect}\label{sec:ecc-mimicker-losa}

\begin{figure*}
  \centering
  \includegraphics[width=\textwidth]{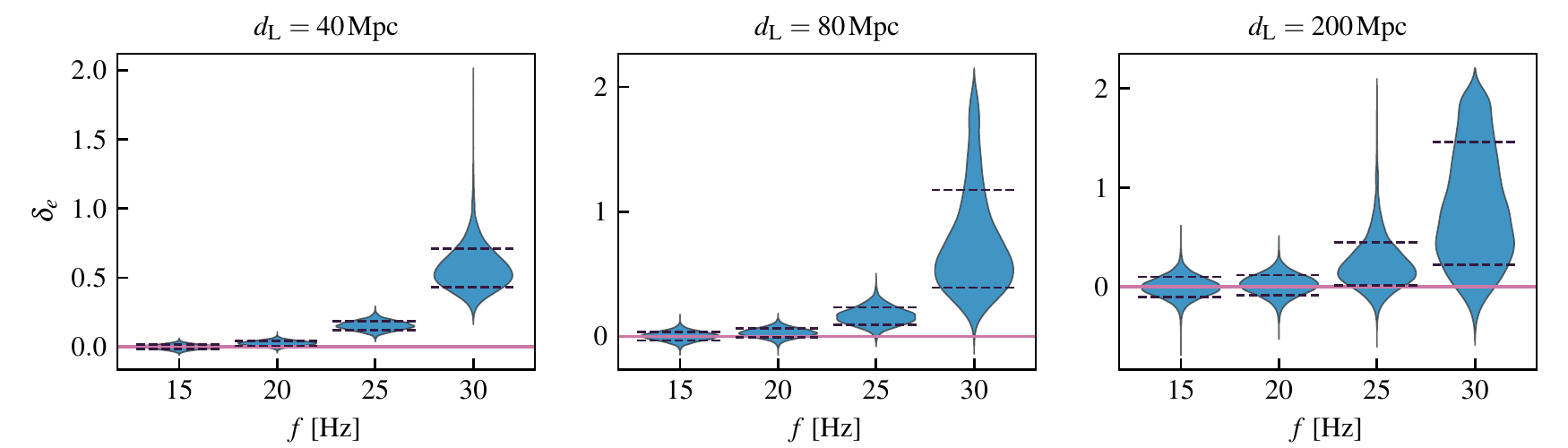}
  \caption{Violins representing eccentricity deviation $\delta_{e}$ are plotted against GW frequency for the case of
    non-spinning quasi-circular zero-noise injection corrected for LOSA effect. CBC
    system has component masses ($5,\,2.5\, M_{\odot}$) with
    center of mass having LOSA parameter $a/c = - 2.25 \times 10^{-4}\, \rm
    s^{-1}$. \textit{Left}, \textit{middle} and \textit{right} panels
    correspond to the systems situated at luminosity distances
    of $40\, \rm Mpc$, $80\, \rm Mpc$ and $200\, \rm Mpc$,
    respectively. In all cases, violation of EECT is clearly observed for $f \geq 25$ Hz, at $\geq 68\%$ confidence.}
  \label{fig: violin_plot_losa}
\end{figure*}
  
Apart from the microlensing effect, we also demonstrate how EECT is able to identify another eccentricity mimicker, i.e., LOSA effect (see Sec.~\ref{sec:losa}). We inject a GW signal representing a quasi-circular CBC with component masses of $m_1= 5\, {M_{\odot}}$ and $m_2= 2.5\, {M_{\odot}}$, and assuming that the center of mass of the binary is having a fiducial value of LOSA $a/c = - 2.25 \times 10^{-4} \rm s^{-1}$. This system is considered at  representative luminosity distances of $d_{\rm L} = 40,\,80,\, 200\,  {\rm Mpc}$. We again recover eccentricity along with other binary parameters at different reference frequencies as mentioned earlier (for the microlensing injection) using {\tt TaylorF2Ecc} as a recovery waveform model. Considering priors on binary parameters as mentioned in Table~\ref{tab:mimicker-prior}, we perform EECT in the same way as already explained in the case of microlensing (cf. Sec.~\ref{sec:ecc-mimicker-mclz}). The SNRs of the recovered signals, in this case,  corresponding to $f_{\rm min}=(10\,{\rm Hz},\,30\, {\rm Hz})$ are given as $(91.89\,,91.75)$, $(47.55\,,45.81)$, and $(18.84\,,18.16)$ for $d_{\rm L} = $ 40, 80, and 200 ${\rm Mpc}$, respectively. The results are shared in Figure~\ref{fig: violin_plot_losa}. We observe that EECT is violated at $68\, \%$ credibility for reference frequencies $f_{i} \ge 25\, {\rm Hz}$ at all three considered luminosity distances. 

\subsubsection{EECT on signal with massive graviton
  effect}\label{sec:ecc-mimicker-MG}
\begin{figure}
    \centering
    \includegraphics[width=\columnwidth]{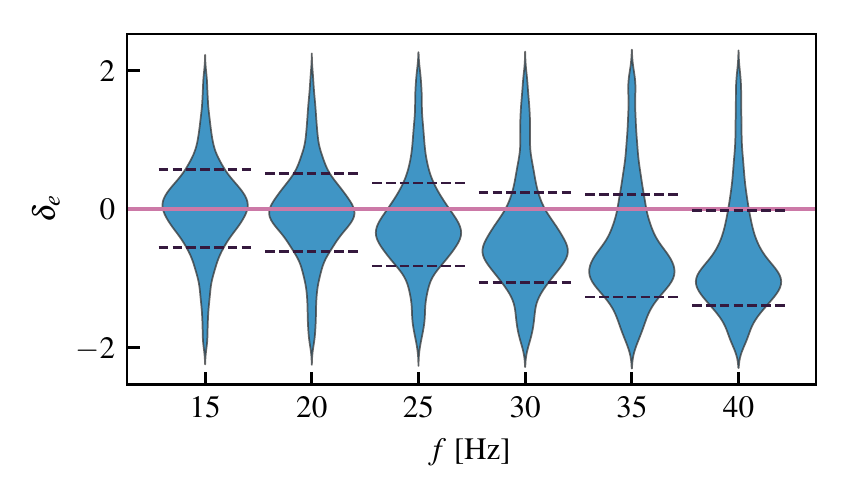}
    \label{fig: MR_diff_post}
    \caption{Violins representing eccentricity deviation $\delta_{e}$ are plotted against GW frequency for the case of non-spinning quasi-circular zero-noise injection corrected for massive graviton effect. CBC has component masses ($60,\,55\, M_{\odot}$) with massive graviton  parameter $A_0 = - 10^{-45}\,  \rm eV^{2}$. The CBC is located at a luminosity distance of $1.8\, \rm Gpc$. At reference frequency $40$Hz, $\delta_e$ is found to be inconsistent with zero at $\geq 68\%$ confidence.}
\end{figure}

For the massive graviton injection, we consider a heavy CBC of component masses ($m_1=60\,M_{\odot},m_2=55\, M_{\odot}$). The modified dispersion effect resulting from the existence of a massive graviton is a propagation effect. Its prominence increases with distance. We therefore consider the CBC at a luminosity distance of $1.8\, {\rm Gpc}$. We set the total mass of the injection to $M= 115\, M_{\odot}$ to ensure sufficient SNR to be detectable, and also enable measurements of non-zero eccentricity with posteriors that are deviated from zero at $\geq 68\%$ confidence. The massive graviton injection parameter $A_0$ is chosen to be $- 10^{-45}\, {\rm eV^2}$ consistent with the bounds obtained by LVK on this parameter~\cite{LIGOScientific:2021sio}. In this case also, the injected signal (using quasi-circular waveform model {\tt TaylorF2} corrected for massive-graviton induced phase correction as described in Sec.~\ref{sec:mg}) is recovered using the eccentric waveform model {\tt TaylorF2Ecc}. Posteriors on eccentricity, along with other parameters, are inferred at different reference frequencies $f_{i}$, and EECT is performed in the same way as in the case of microlensing and LOSA effects. Priors on the binary parameters are mentioned in Table~\ref{tab:mimicker-prior}. The SNR of the recovered signals, in this case,  corresponding to $f_{\rm min}=(10\,{\rm Hz},\,40\, {\rm Hz})$ are given as $(20.60,\, 18.66)$. 
Figure~\ref{fig: MR_diff_post} shows how the posterior on $\delta_{e}$ varies as a function of GW reference frequency. We observe that the EECT is violated at $f_{i} = 40 {\rm Hz}$.

\subsubsection{EECT on signal with dipole radiation effect}\label{sec:ecc-mimicker-DR}

\begin{figure*}
    \centering
    \includegraphics[width=\textwidth]{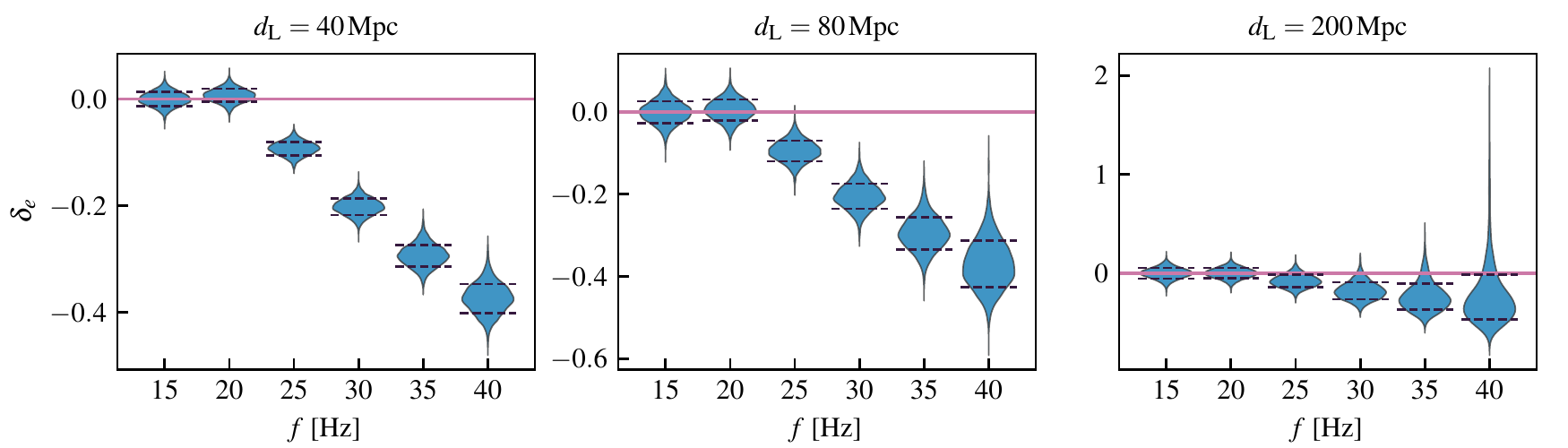}
    \label{fig: DR_diff_post}
    \caption{Violins representing eccentricity deviation $\delta_{\rm
        e}$ are plotted against GW frequency for a non-spinning
      quasi-circular zero-noise injection corrected for dipole radiation effect. CBC system has component masses ($5,\,2.5\,
      M_{\odot}$) with dipole radiation parameter $B = 1.24 \times
      10^{-3}$. \textit{Left}, \textit{middle} and
      \textit{right} panels correspond to the system being considered
      at luminosity distances of $40\, \rm Mpc$, $80\, \rm Mpc$ and
      $200\, \rm Mpc$, respectively. In all cases, EECT is found to be violated at $\geq 68\%$ confidence for $f \geq 25$ Hz.}
\end{figure*}

For dipole radiation effect operating in the otherwise quasi-circular
waveform model {\tt TaylorF2}, EECT is performed in the same way as
demonstrated already for other mimicker effects. Note that the priors considered while recovering the binary parameters
of the injected GW signal are mentioned in Table~\ref{tab:mimicker-prior}. The dipole radiation parameter $B$
appearing in the $-1$PN correction to the quasi-circular model {\tt
  TaylorF2} as described in Sec.~\ref{sec:dipole} is chosen to be
$1.24 \times 10^{-3}$ consistent with the LVK bound
obtained on this parameter~\cite{LIGOScientific:2021sio}. The mass of the binary system is
chosen to be $7.5\, {M_{\odot}}$ (with mass ratio $q=0.5$) and the binary system is considered at representative luminosity distances of $d_{\rm L}= 40\,,80\,,200\, {\rm Mpc}$. The SNRs of the recovered signals, in this case,  corresponding to $f_{\rm min}=(10\,{\rm Hz},\,40\, {\rm Hz})$ are given as $(88.87\,,82.17)$, $(44.31\,,40.96)$, and $(17.56\,,16.15)$ for $d_{\rm L} = $ 40, 80, and 200 ${\rm Mpc}$, respectively. 
Figure~\ref{fig: DR_diff_post} displays the variation of the posterior on the eccentricity deviation
parameter $\delta_{e}$ with reference frequency. We observe
that EECT is violated at $f_{i} \ge 25 {\rm Hz}$
for all the considered distances.

\section{Summary and Outlook}\label{sec:conclusion}
The number of CBC events, detected by the LVK, for which signatures of eccentricity have been reported, has been steadily increasing \cite{RomeroShaw2022_multi_ecc, RomeroShaw2020_GW190521_eccentric, Romero-Shaw:2025vbc, Morras2025_GW200105_ecc, Planas2025_GW200105_IMR_ecc, Gupte2024_pop_ecc, DhurkundeNitz2025_eccSearch_O3, Kacanja:2025kpr}. Given the importance of residual eccentricity measurements in GWs in the context of probing CBC formation channels and testing GR, it is crucial to ascertain that what is being measured is indeed an eccentric signal, and not an eccentricity mimicker. For example, GW190521, the heaviest BBH merger of the first three observing runs of the LVK, has been claimed to be eccentric \cite{RomeroShaw2020_GW190521_eccentric, Gayathri2020_GW190521_NR}, and precessing \cite{Miller2023_GW190521_precessing}. It has also been suggested that this merger could be a head-on collision instead of a precessing one \cite{CalderonBustillo2020_headOnGW190521}, and even a merger of proca-stars \cite{CalderonBustillo2020_ProcaHeadOn}.

Currently,
Bayesian model comparison is the standard method employed to identify
eccentricity among a plethora of other possible hypotheses that could mimic eccentricity, such as precession \cite{RomeroShaw2023_precession_eccentricity_degeneracy}, microlensing \cite{mishra-ecc-mclz}, and LOSA \cite{Tiwari:2025aec}. This
method is not only computationally expensive but might be misleading if none of the models under consideration capture the true physical or beyond-GR effect modulating the GWs. {Moreover, it should be noted that certain physical effects, such as multiple microlenses in a macro-potential, are unfeasible to probe with PE because generating those models is computationally expensive~\cite{2021MNRAS.508.4869M}.} This motivates us to propose a novel, simple, and powerful
method to identify eccentric signals and reject mimickers, {\it without needing to invoke other models}. The method compares eccentricity recovered at different reference frequencies with those predicted by evolving eccentricity measured at some fiducial value, and assuming that the CBC undergoes GR-consistent isolated evolution. The underlying expectation is that while spurious eccentricity values at different reference frequencies may be produced when recovering a mimicker with eccentric GW waveform models, the frequency evolution of eccentricity will in general not be mimicked. 

As a proof of principle, we first consider three eccentric non-spinning CBCs with distinct total masses (light, medium and heavy), with finite eccentricities at $10$ Hz in a 3-detector O4-like scenario, and show that EECT is satisfied at $\geq 68\%$ confidence for these systems. We then consider four non-spinning eccentricity mimicker models, two quasi-circular waveforms modulated by physical effects (microlensing and LOSA), and two modulated by beyond-GR effects (dipole radiation and massive graviton). We recover these with non-spinning eccentric waveform models. We find that, for the systems we consider, EECT is indeed violated at $\geq 68\%$ confidence for frequencies above some reference frequency that depends on the parameters of the mimicker, demonstrating the power of the method. 

{The method crucially relies on the precision with which eccentricities can be recovered, which in turn depends on the sensitivity and bandwidth of the detector. Eccentricities with $e \lesssim 0.01$, for the O4-like sensitivity we consider, result in significant railing of the posterior against the prior ``wall'' at $e = 0$. Such railing could result in spurious violations of EECT, and frequencies above which such railing is extensive should be rejected. In our work, we ensure that $e=0$ is excluded at $\geq 68\%$ confidence for reference frequencies at which $\delta_e$ is evaluated. This, however, ultimately defines the CBC parameters over which EECT may be applied, because there are systems -- those with large total masses (and therefore fewer GW cycles in-band), and/or low SNR -- where said $e=0$ exclusion criterion is not satisfied. Nevertheless, improved sensitivity and larger frequency bands, such as those planned in future ground \cite{Reitze:2019iox, punturo2010} and space-based \cite{LISA:2017pwj, Sato:2017dkf} detectors, should enable a significantly larger CBC parameter space over which EECT could be applied.} 

Our proof-of-concept work demonstrates the power and promise of EECT. Nevertheless, it is also of interest to clearly demarcate the region of parameter space over which EECT is applicable, as a function of sensitivities associated with various observing scenarios and detector networks. We plan to do so in future work, where we also intend to use state of the art waveforms that incorporate the merger-ringdown phase, higher harmonics, precessing spins, and mean anomaly. {{Moreover, we keep the detailed study on the application of EECT in distinguishing spin-precession effect from eccentricity as a future follow up work. }}

\begin{acknowledgements}
  The authors would like to thank Mukesh Kumar Singh for valuable
  feedback on the draft. S. A. B. acknowledges useful comments by the participants of GR24 and Amaldi16 joint conference during the presentation of this work. M. A. S. acknowledges hospitality at IUCAA during the visit related to this work. S.J.K. acknowledges support from SERB grants SRG/2023/000419 and MTR/2023/000086. We gratefully acknowledge computational resources provided by the LIGO Laboratory, a major facility fully funded by the National Science Foundation under Grants PHY-0757058 and PHY-0823459.
\end{acknowledgements}

\bibliography{References}

\begin{thebibliography}{86}%
\makeatletter
\providecommand \@ifxundefined [1]{%
 \@ifx{#1\undefined}
}%
\providecommand \@ifnum [1]{%
 \ifnum #1\expandafter \@firstoftwo
 \else \expandafter \@secondoftwo
 \fi
}%
\providecommand \@ifx [1]{%
 \ifx #1\expandafter \@firstoftwo
 \else \expandafter \@secondoftwo
 \fi
}%
\providecommand \natexlab [1]{#1}%
\providecommand \enquote  [1]{``#1''}%
\providecommand \bibnamefont  [1]{#1}%
\providecommand \bibfnamefont [1]{#1}%
\providecommand \citenamefont [1]{#1}%
\providecommand \href@noop [0]{\@secondoftwo}%
\providecommand \href [0]{\begingroup \@sanitize@url \@href}%
\providecommand \@href[1]{\@@startlink{#1}\@@href}%
\providecommand \@@href[1]{\endgroup#1\@@endlink}%
\providecommand \@sanitize@url [0]{\catcode `\\12\catcode `\$12\catcode
  `\&12\catcode `\#12\catcode `\^12\catcode `\_12\catcode `\%12\relax}%
\providecommand \@@startlink[1]{}%
\providecommand \@@endlink[0]{}%
\providecommand \url  [0]{\begingroup\@sanitize@url \@url }%
\providecommand \@url [1]{\endgroup\@href {#1}{\urlprefix }}%
\providecommand \urlprefix  [0]{URL }%
\providecommand \Eprint [0]{\href }%
\providecommand \doibase [0]{http://dx.doi.org/}%
\providecommand \selectlanguage [0]{\@gobble}%
\providecommand \bibinfo  [0]{\@secondoftwo}%
\providecommand \bibfield  [0]{\@secondoftwo}%
\providecommand \translation [1]{[#1]}%
\providecommand \BibitemOpen [0]{}%
\providecommand \bibitemStop [0]{}%
\providecommand \bibitemNoStop [0]{.\EOS\space}%
\providecommand \EOS [0]{\spacefactor3000\relax}%
\providecommand \BibitemShut  [1]{\csname bibitem#1\endcsname}%
\let\auto@bib@innerbib\@empty
\bibitem [{\citenamefont {Aasi}\ \emph {et~al.}(2015)\citenamefont {Aasi} \emph
  {et~al.}}]{TheLIGOScientific:2014jea}%
  \BibitemOpen
  \bibfield  {author} {\bibinfo {author} {\bibfnamefont {J.}~\bibnamefont
  {Aasi}} \emph {et~al.} (\bibinfo {collaboration} {LIGO Scientific}),\
  }\bibfield  {title} {\enquote {\bibinfo {title} {{Advanced LIGO}},}\ }\href
  {\doibase 10.1088/0264-9381/32/7/074001} {\bibfield  {journal} {\bibinfo
  {journal} {Class. Quant. Grav.}\ }\textbf {\bibinfo {volume} {32}},\ \bibinfo
  {pages} {074001} (\bibinfo {year} {2015})},\ \Eprint
  {http://arxiv.org/abs/1411.4547} {arXiv:1411.4547 [gr-qc]} \BibitemShut
  {NoStop}%
\bibitem [{\citenamefont {Acernese}\ \emph {et~al.}(2015)\citenamefont
  {Acernese} \emph {et~al.}}]{TheVirgo:2014hva}%
  \BibitemOpen
  \bibfield  {author} {\bibinfo {author} {\bibfnamefont {F.}~\bibnamefont
  {Acernese}} \emph {et~al.} (\bibinfo {collaboration} {Virgo}),\ }\bibfield
  {title} {\enquote {\bibinfo {title} {{Advanced Virgo: a second-generation
  interferometric gravitational wave detector}},}\ }\href {\doibase
  10.1088/0264-9381/32/2/024001} {\bibfield  {journal} {\bibinfo  {journal}
  {Class. Quant. Grav.}\ }\textbf {\bibinfo {volume} {32}},\ \bibinfo {pages}
  {024001} (\bibinfo {year} {2015})},\ \Eprint {http://arxiv.org/abs/1408.3978}
  {arXiv:1408.3978 [gr-qc]} \BibitemShut {NoStop}%
\bibitem [{\citenamefont {Akutsu}\ \emph {et~al.}(2021)\citenamefont {Akutsu}
  \emph {et~al.}}]{KAGRA:2020tym}%
  \BibitemOpen
  \bibfield  {author} {\bibinfo {author} {\bibfnamefont {T.}~\bibnamefont
  {Akutsu}} \emph {et~al.} (\bibinfo {collaboration} {KAGRA}),\ }\bibfield
  {title} {\enquote {\bibinfo {title} {{Overview of KAGRA: Detector design and
  construction history}},}\ }\href {\doibase 10.1093/ptep/ptaa125} {\bibfield
  {journal} {\bibinfo  {journal} {PTEP}\ }\textbf {\bibinfo {volume} {2021}},\
  \bibinfo {pages} {05A101} (\bibinfo {year} {2021})},\ \Eprint
  {http://arxiv.org/abs/2005.05574} {arXiv:2005.05574 [physics.ins-det]}
  \BibitemShut {NoStop}%
\bibitem [{\citenamefont {Abbott}\ \emph
  {et~al.}(2019{\natexlab{a}})\citenamefont {Abbott} \emph
  {et~al.}}]{LIGOScientific:2018mvr}%
  \BibitemOpen
  \bibfield  {author} {\bibinfo {author} {\bibfnamefont {B.~P.}\ \bibnamefont
  {Abbott}} \emph {et~al.} (\bibinfo {collaboration} {LIGO Scientific,
  Virgo}),\ }\bibfield  {title} {\enquote {\bibinfo {title} {{GWTC-1: A
  Gravitational-Wave Transient Catalog of Compact Binary Mergers Observed by
  LIGO and Virgo during the First and Second Observing Runs}},}\ }\href
  {\doibase 10.1103/PhysRevX.9.031040} {\bibfield  {journal} {\bibinfo
  {journal} {Phys. Rev. X}\ }\textbf {\bibinfo {volume} {9}},\ \bibinfo {pages}
  {031040} (\bibinfo {year} {2019}{\natexlab{a}})},\ \Eprint
  {http://arxiv.org/abs/1811.12907} {arXiv:1811.12907 [astro-ph.HE]}
  \BibitemShut {NoStop}%
\bibitem [{\citenamefont {Abbott}\ \emph
  {et~al.}(2021{\natexlab{a}})\citenamefont {Abbott} \emph
  {et~al.}}]{LIGOScientific:2020ibl}%
  \BibitemOpen
  \bibfield  {author} {\bibinfo {author} {\bibfnamefont {R.}~\bibnamefont
  {Abbott}} \emph {et~al.} (\bibinfo {collaboration} {LIGO Scientific,
  Virgo}),\ }\bibfield  {title} {\enquote {\bibinfo {title} {{GWTC-2: Compact
  Binary Coalescences Observed by LIGO and Virgo During the First Half of the
  Third Observing Run}},}\ }\href {\doibase 10.1103/PhysRevX.11.021053}
  {\bibfield  {journal} {\bibinfo  {journal} {Phys. Rev. X}\ }\textbf {\bibinfo
  {volume} {11}},\ \bibinfo {pages} {021053} (\bibinfo {year}
  {2021}{\natexlab{a}})},\ \Eprint {http://arxiv.org/abs/2010.14527}
  {arXiv:2010.14527 [gr-qc]} \BibitemShut {NoStop}%
\bibitem [{\citenamefont {Abbott}\ \emph {et~al.}(2023)\citenamefont {Abbott}
  \emph {et~al.}}]{KAGRA:2021vkt}%
  \BibitemOpen
  \bibfield  {author} {\bibinfo {author} {\bibfnamefont {R.}~\bibnamefont
  {Abbott}} \emph {et~al.} (\bibinfo {collaboration} {KAGRA, VIRGO, LIGO
  Scientific}),\ }\bibfield  {title} {\enquote {\bibinfo {title} {{GWTC-3:
  Compact Binary Coalescences Observed by LIGO and Virgo during the Second Part
  of the Third Observing Run}},}\ }\href {\doibase 10.1103/PhysRevX.13.041039}
  {\bibfield  {journal} {\bibinfo  {journal} {Phys. Rev. X}\ }\textbf {\bibinfo
  {volume} {13}},\ \bibinfo {pages} {041039} (\bibinfo {year} {2023})},\
  \Eprint {http://arxiv.org/abs/2111.03606} {arXiv:2111.03606 [gr-qc]}
  \BibitemShut {NoStop}%
\bibitem [{\citenamefont {Abac}\ \emph
  {et~al.}(2025{\natexlab{a}})\citenamefont {Abac} \emph
  {et~al.}}]{LIGOScientific:2025slb}%
  \BibitemOpen
  \bibfield  {author} {\bibinfo {author} {\bibfnamefont {A.~G.}\ \bibnamefont
  {Abac}} \emph {et~al.} (\bibinfo {collaboration} {LIGO Scientific, VIRGO,
  KAGRA}),\ }\bibfield  {title} {\enquote {\bibinfo {title} {{GWTC-4.0:
  Updating the Gravitational-Wave Transient Catalog with Observations from the
  First Part of the Fourth LIGO-Virgo-KAGRA Observing Run}},}\ }\href@noop {}
  {\  (\bibinfo {year} {2025}{\natexlab{a}})},\ \Eprint
  {http://arxiv.org/abs/2508.18082} {arXiv:2508.18082 [gr-qc]} \BibitemShut
  {NoStop}%
\bibitem [{\citenamefont {Abac}\ \emph
  {et~al.}(2025{\natexlab{b}})\citenamefont {Abac} \emph
  {et~al.}}]{LIGOScientific:2025hdt}%
  \BibitemOpen
  \bibfield  {author} {\bibinfo {author} {\bibfnamefont {A.~G.}\ \bibnamefont
  {Abac}} \emph {et~al.} (\bibinfo {collaboration} {LIGO Scientific, VIRGO,
  KAGRA}),\ }\bibfield  {title} {\enquote {\bibinfo {title} {{GWTC-4.0: An
  Introduction to Version 4.0 of the Gravitational-Wave Transient Catalog}},}\
  }\href@noop {} {\  (\bibinfo {year} {2025}{\natexlab{b}})},\ \Eprint
  {http://arxiv.org/abs/2508.18080} {arXiv:2508.18080 [gr-qc]} \BibitemShut
  {NoStop}%
\bibitem [{\citenamefont {Abac}\ \emph
  {et~al.}(2025{\natexlab{c}})\citenamefont {Abac} \emph
  {et~al.}}]{LIGOScientific:2025yae}%
  \BibitemOpen
  \bibfield  {author} {\bibinfo {author} {\bibfnamefont {A.~G.}\ \bibnamefont
  {Abac}} \emph {et~al.} (\bibinfo {collaboration} {LIGO Scientific, VIRGO,
  KAGRA}),\ }\bibfield  {title} {\enquote {\bibinfo {title} {{GWTC-4.0: Methods
  for Identifying and Characterizing Gravitational-wave Transients}},}\
  }\href@noop {} {\  (\bibinfo {year} {2025}{\natexlab{c}})},\ \Eprint
  {http://arxiv.org/abs/2508.18081} {arXiv:2508.18081 [gr-qc]} \BibitemShut
  {NoStop}%
\bibitem [{\citenamefont {Abbott}\ \emph {et~al.}(2017)\citenamefont {Abbott}
  \emph {et~al.}}]{TheLIGOScientific:2017qsa}%
  \BibitemOpen
  \bibfield  {author} {\bibinfo {author} {\bibfnamefont {Benjamin~P.}\
  \bibnamefont {Abbott}} \emph {et~al.} (\bibinfo {collaboration} {LIGO
  Scientific, Virgo}),\ }\bibfield  {title} {\enquote {\bibinfo {title}
  {{GW170817: Observation of Gravitational Waves from a Binary Neutron Star
  Inspiral}},}\ }\href {\doibase 10.1103/PhysRevLett.119.161101} {\bibfield
  {journal} {\bibinfo  {journal} {Phys. Rev. Lett.}\ }\textbf {\bibinfo
  {volume} {119}},\ \bibinfo {pages} {161101} (\bibinfo {year} {2017})},\
  \Eprint {http://arxiv.org/abs/1710.05832} {arXiv:1710.05832 [gr-qc]}
  \BibitemShut {NoStop}%
\bibitem [{\citenamefont {Abbott}\ \emph {et~al.}(2020)\citenamefont {Abbott}
  \emph {et~al.}}]{Abbott:2020uma}%
  \BibitemOpen
  \bibfield  {author} {\bibinfo {author} {\bibfnamefont {B.P.}\ \bibnamefont
  {Abbott}} \emph {et~al.} (\bibinfo {collaboration} {LIGO Scientific,
  Virgo}),\ }\bibfield  {title} {\enquote {\bibinfo {title} {{GW190425:
  Observation of a Compact Binary Coalescence with Total Mass $\sim 3.4
  M_{\odot}$}},}\ }\href {\doibase 10.3847/2041-8213/ab75f5} {\bibfield
  {journal} {\bibinfo  {journal} {Astrophys. J. Lett.}\ }\textbf {\bibinfo
  {volume} {892}},\ \bibinfo {pages} {L3} (\bibinfo {year} {2020})},\ \Eprint
  {http://arxiv.org/abs/2001.01761} {arXiv:2001.01761 [astro-ph.HE]}
  \BibitemShut {NoStop}%
\bibitem [{\citenamefont {Abbott}\ \emph
  {et~al.}(2021{\natexlab{b}})\citenamefont {Abbott} \emph
  {et~al.}}]{LIGOScientific:2021qlt}%
  \BibitemOpen
  \bibfield  {author} {\bibinfo {author} {\bibfnamefont {R.}~\bibnamefont
  {Abbott}} \emph {et~al.} (\bibinfo {collaboration} {LIGO Scientific, KAGRA,
  VIRGO}),\ }\bibfield  {title} {\enquote {\bibinfo {title} {{Observation of
  Gravitational Waves from Two Neutron Star\textendash{}Black Hole
  Coalescences}},}\ }\href {\doibase 10.3847/2041-8213/ac082e} {\bibfield
  {journal} {\bibinfo  {journal} {Astrophys. J. Lett.}\ }\textbf {\bibinfo
  {volume} {915}},\ \bibinfo {pages} {L5} (\bibinfo {year}
  {2021}{\natexlab{b}})},\ \Eprint {http://arxiv.org/abs/2106.15163}
  {arXiv:2106.15163 [astro-ph.HE]} \BibitemShut {NoStop}%
\bibitem [{\citenamefont {Abbott}\ \emph
  {et~al.}(2021{\natexlab{c}})\citenamefont {Abbott} \emph
  {et~al.}}]{LIGOScientific:2021psn}%
  \BibitemOpen
  \bibfield  {author} {\bibinfo {author} {\bibfnamefont {R.}~\bibnamefont
  {Abbott}} \emph {et~al.} (\bibinfo {collaboration} {LIGO Scientific, VIRGO,
  KAGRA}),\ }\bibfield  {title} {\enquote {\bibinfo {title} {{The population of
  merging compact binaries inferred using gravitational waves through
  GWTC-3}},}\ }\href@noop {} {\  (\bibinfo {year} {2021}{\natexlab{c}})},\
  \Eprint {http://arxiv.org/abs/2111.03634} {arXiv:2111.03634 [astro-ph.HE]}
  \BibitemShut {NoStop}%
\bibitem [{\citenamefont {Zevin}\ \emph {et~al.}(2021)\citenamefont {Zevin},
  \citenamefont {Bavera}, \citenamefont {Berry}, \citenamefont {Kalogera},
  \citenamefont {Fragos}, \citenamefont {Marchant}, \citenamefont {Rodriguez},
  \citenamefont {Antonini}, \citenamefont {Holz},\ and\ \citenamefont
  {Pankow}}]{Zevin:2020gbd}%
  \BibitemOpen
  \bibfield  {author} {\bibinfo {author} {\bibfnamefont {Michael}\ \bibnamefont
  {Zevin}}, \bibinfo {author} {\bibfnamefont {Simone~S.}\ \bibnamefont
  {Bavera}}, \bibinfo {author} {\bibfnamefont {Christopher P.~L.}\ \bibnamefont
  {Berry}}, \bibinfo {author} {\bibfnamefont {Vicky}\ \bibnamefont {Kalogera}},
  \bibinfo {author} {\bibfnamefont {Tassos}\ \bibnamefont {Fragos}}, \bibinfo
  {author} {\bibfnamefont {Pablo}\ \bibnamefont {Marchant}}, \bibinfo {author}
  {\bibfnamefont {Carl~L.}\ \bibnamefont {Rodriguez}}, \bibinfo {author}
  {\bibfnamefont {Fabio}\ \bibnamefont {Antonini}}, \bibinfo {author}
  {\bibfnamefont {Daniel~E.}\ \bibnamefont {Holz}}, \ and\ \bibinfo {author}
  {\bibfnamefont {Chris}\ \bibnamefont {Pankow}},\ }\bibfield  {title}
  {\enquote {\bibinfo {title} {{One Channel to Rule Them All? Constraining the
  Origins of Binary Black Holes Using Multiple Formation Pathways}},}\ }\href
  {\doibase 10.3847/1538-4357/abe40e} {\bibfield  {journal} {\bibinfo
  {journal} {Astrophys. J.}\ }\textbf {\bibinfo {volume} {910}},\ \bibinfo
  {pages} {152} (\bibinfo {year} {2021})},\ \Eprint
  {http://arxiv.org/abs/2011.10057} {arXiv:2011.10057 [astro-ph.HE]}
  \BibitemShut {NoStop}%
\bibitem [{\citenamefont {Zevin}\ \emph {et~al.}(2017)\citenamefont {Zevin},
  \citenamefont {Pankow}, \citenamefont {Rodriguez}, \citenamefont {Sampson},
  \citenamefont {Chase}, \citenamefont {Kalogera},\ and\ \citenamefont
  {Rasio}}]{Zevin:2017evb}%
  \BibitemOpen
  \bibfield  {author} {\bibinfo {author} {\bibfnamefont {Michael}\ \bibnamefont
  {Zevin}}, \bibinfo {author} {\bibfnamefont {Chris}\ \bibnamefont {Pankow}},
  \bibinfo {author} {\bibfnamefont {Carl~L.}\ \bibnamefont {Rodriguez}},
  \bibinfo {author} {\bibfnamefont {Laura}\ \bibnamefont {Sampson}}, \bibinfo
  {author} {\bibfnamefont {Eve}\ \bibnamefont {Chase}}, \bibinfo {author}
  {\bibfnamefont {Vassiliki}\ \bibnamefont {Kalogera}}, \ and\ \bibinfo
  {author} {\bibfnamefont {Frederic~A.}\ \bibnamefont {Rasio}},\ }\bibfield
  {title} {\enquote {\bibinfo {title} {{Constraining Formation Models of Binary
  Black Holes with Gravitational-Wave Observations}},}\ }\href {\doibase
  10.3847/1538-4357/aa8408} {\bibfield  {journal} {\bibinfo  {journal}
  {Astrophys. J.}\ }\textbf {\bibinfo {volume} {846}},\ \bibinfo {pages} {82}
  (\bibinfo {year} {2017})},\ \Eprint {http://arxiv.org/abs/1704.07379}
  {arXiv:1704.07379 [astro-ph.HE]} \BibitemShut {NoStop}%
\bibitem [{\citenamefont {Mandel}\ and\ \citenamefont
  {Farmer}(2022)}]{Mandel:2018hfr}%
  \BibitemOpen
  \bibfield  {author} {\bibinfo {author} {\bibfnamefont {Ilya}\ \bibnamefont
  {Mandel}}\ and\ \bibinfo {author} {\bibfnamefont {Alison}\ \bibnamefont
  {Farmer}},\ }\bibfield  {title} {\enquote {\bibinfo {title} {{Merging
  stellar-mass binary black holes}},}\ }\href {\doibase
  10.1016/j.physrep.2022.01.003} {\bibfield  {journal} {\bibinfo  {journal}
  {Phys. Rept.}\ }\textbf {\bibinfo {volume} {955}},\ \bibinfo {pages} {1--24}
  (\bibinfo {year} {2022})},\ \Eprint {http://arxiv.org/abs/1806.05820}
  {arXiv:1806.05820 [astro-ph.HE]} \BibitemShut {NoStop}%
\bibitem [{\citenamefont {Taylor}\ and\ \citenamefont
  {Gerosa}(2018)}]{Taylor:2018iat}%
  \BibitemOpen
  \bibfield  {author} {\bibinfo {author} {\bibfnamefont {Stephen~R.}\
  \bibnamefont {Taylor}}\ and\ \bibinfo {author} {\bibfnamefont {Davide}\
  \bibnamefont {Gerosa}},\ }\bibfield  {title} {\enquote {\bibinfo {title}
  {{Mining Gravitational-wave Catalogs To Understand Binary Stellar Evolution:
  A New Hierarchical Bayesian Framework}},}\ }\href {\doibase
  10.1103/PhysRevD.98.083017} {\bibfield  {journal} {\bibinfo  {journal} {Phys.
  Rev. D}\ }\textbf {\bibinfo {volume} {98}},\ \bibinfo {pages} {083017}
  (\bibinfo {year} {2018})},\ \Eprint {http://arxiv.org/abs/1806.08365}
  {arXiv:1806.08365 [astro-ph.HE]} \BibitemShut {NoStop}%
\bibitem [{\citenamefont {Roulet}\ and\ \citenamefont
  {Zaldarriaga}(2019)}]{Roulet:2018jbe}%
  \BibitemOpen
  \bibfield  {author} {\bibinfo {author} {\bibfnamefont {Javier}\ \bibnamefont
  {Roulet}}\ and\ \bibinfo {author} {\bibfnamefont {Matias}\ \bibnamefont
  {Zaldarriaga}},\ }\bibfield  {title} {\enquote {\bibinfo {title}
  {{Constraints on binary black hole populations from LIGO{\textendash}Virgo
  detections}},}\ }\href {\doibase 10.1093/mnras/stz226} {\bibfield  {journal}
  {\bibinfo  {journal} {Mon. Not. Roy. Astron. Soc.}\ }\textbf {\bibinfo
  {volume} {484}},\ \bibinfo {pages} {4216--4229} (\bibinfo {year} {2019})},\
  \Eprint {http://arxiv.org/abs/1806.10610} {arXiv:1806.10610 [astro-ph.HE]}
  \BibitemShut {NoStop}%
\bibitem [{\citenamefont {Baibhav}\ \emph {et~al.}(2020)\citenamefont
  {Baibhav}, \citenamefont {Gerosa}, \citenamefont {Berti}, \citenamefont
  {Wong}, \citenamefont {Helfer},\ and\ \citenamefont
  {Mould}}]{Baibhav:2020xdf}%
  \BibitemOpen
  \bibfield  {author} {\bibinfo {author} {\bibfnamefont {Vishal}\ \bibnamefont
  {Baibhav}}, \bibinfo {author} {\bibfnamefont {Davide}\ \bibnamefont
  {Gerosa}}, \bibinfo {author} {\bibfnamefont {Emanuele}\ \bibnamefont
  {Berti}}, \bibinfo {author} {\bibfnamefont {Kaze W.~K.}\ \bibnamefont
  {Wong}}, \bibinfo {author} {\bibfnamefont {Thomas}\ \bibnamefont {Helfer}}, \
  and\ \bibinfo {author} {\bibfnamefont {Matthew}\ \bibnamefont {Mould}},\
  }\bibfield  {title} {\enquote {\bibinfo {title} {{The mass gap, the spin gap,
  and the origin of merging binary black holes}},}\ }\href {\doibase
  10.1103/PhysRevD.102.043002} {\bibfield  {journal} {\bibinfo  {journal}
  {Phys. Rev. D}\ }\textbf {\bibinfo {volume} {102}},\ \bibinfo {pages}
  {043002} (\bibinfo {year} {2020})},\ \Eprint
  {http://arxiv.org/abs/2004.00650} {arXiv:2004.00650 [astro-ph.HE]}
  \BibitemShut {NoStop}%
\bibitem [{\citenamefont {{Mapelli}}(2020)}]{Mapelli:2021for}%
  \BibitemOpen
  \bibfield  {author} {\bibinfo {author} {\bibfnamefont {Michela}\ \bibnamefont
  {{Mapelli}}},\ }\bibfield  {title} {\enquote {\bibinfo {title} {{Binary black
  hole mergers: formation and populations}},}\ }\href {\doibase
  10.3389/fspas.2020.00038} {\bibfield  {journal} {\bibinfo  {journal}
  {Frontiers in Astronomy and Space Sciences}\ }\textbf {\bibinfo {volume}
  {7}},\ \bibinfo {eid} {38} (\bibinfo {year} {2020})},\ \Eprint
  {http://arxiv.org/abs/2105.12455} {arXiv:2105.12455 [astro-ph.HE]}
  \BibitemShut {NoStop}%
\bibitem [{\citenamefont {Afroz}\ and\ \citenamefont
  {Mukherjee}(2025)}]{Afroz:2024fzp}%
  \BibitemOpen
  \bibfield  {author} {\bibinfo {author} {\bibfnamefont {Samsuzzaman}\
  \bibnamefont {Afroz}}\ and\ \bibinfo {author} {\bibfnamefont {Suvodip}\
  \bibnamefont {Mukherjee}},\ }\bibfield  {title} {\enquote {\bibinfo {title}
  {{Phase space of binary black holes from gravitational wave observations to
  unveil its formation history}},}\ }\href {\doibase 10.1103/7zc2-g9vq}
  {\bibfield  {journal} {\bibinfo  {journal} {Phys. Rev. D}\ }\textbf {\bibinfo
  {volume} {112}},\ \bibinfo {pages} {023531} (\bibinfo {year} {2025})},\
  \Eprint {http://arxiv.org/abs/2411.07304} {arXiv:2411.07304 [astro-ph.HE]}
  \BibitemShut {NoStop}%
\bibitem [{\citenamefont {Stegmann}\ \emph {et~al.}(2025)\citenamefont
  {Stegmann}, \citenamefont {Gerosa}, \citenamefont {Romero-Shaw},
  \citenamefont {Fumagalli}, \citenamefont {Tagawa},\ and\ \citenamefont
  {Zwick}}]{Stegmann:2025shr}%
  \BibitemOpen
  \bibfield  {author} {\bibinfo {author} {\bibfnamefont {Jakob}\ \bibnamefont
  {Stegmann}}, \bibinfo {author} {\bibfnamefont {Davide}\ \bibnamefont
  {Gerosa}}, \bibinfo {author} {\bibfnamefont {Isobel}\ \bibnamefont
  {Romero-Shaw}}, \bibinfo {author} {\bibfnamefont {Giulia}\ \bibnamefont
  {Fumagalli}}, \bibinfo {author} {\bibfnamefont {Hiromichi}\ \bibnamefont
  {Tagawa}}, \ and\ \bibinfo {author} {\bibfnamefont {Lorenz}\ \bibnamefont
  {Zwick}},\ }\bibfield  {title} {\enquote {\bibinfo {title} {{Distinguishing
  the origin of eccentric black-hole mergers with gravitational-wave spin
  measurements}},}\ }\href@noop {} {\  (\bibinfo {year} {2025})},\ \Eprint
  {http://arxiv.org/abs/2505.13589} {arXiv:2505.13589 [astro-ph.HE]}
  \BibitemShut {NoStop}%
\bibitem [{\citenamefont {Dorozsmai}\ \emph {et~al.}(2025)\citenamefont
  {Dorozsmai}, \citenamefont {Romero-Shaw}, \citenamefont {Vijaykumar},
  \citenamefont {Toonen}, \citenamefont {Antonini}, \citenamefont {Kremer},
  \citenamefont {Zevin},\ and\ \citenamefont {Grishin}}]{Dorozsmai:2025jlu}%
  \BibitemOpen
  \bibfield  {author} {\bibinfo {author} {\bibfnamefont {Andris}\ \bibnamefont
  {Dorozsmai}}, \bibinfo {author} {\bibfnamefont {Isobel~M.}\ \bibnamefont
  {Romero-Shaw}}, \bibinfo {author} {\bibfnamefont {Aditya}\ \bibnamefont
  {Vijaykumar}}, \bibinfo {author} {\bibfnamefont {Silvia}\ \bibnamefont
  {Toonen}}, \bibinfo {author} {\bibfnamefont {Fabio}\ \bibnamefont
  {Antonini}}, \bibinfo {author} {\bibfnamefont {Kyle}\ \bibnamefont {Kremer}},
  \bibinfo {author} {\bibfnamefont {Michael}\ \bibnamefont {Zevin}}, \ and\
  \bibinfo {author} {\bibfnamefont {Evgeni}\ \bibnamefont {Grishin}},\
  }\bibfield  {title} {\enquote {\bibinfo {title} {{Hierarchical Triples vs.
  Globular Clusters: Binary black hole merger eccentricity distributions
  compete and evolve with redshift}},}\ }\href@noop {} {\  (\bibinfo {year}
  {2025})},\ \Eprint {http://arxiv.org/abs/2507.23212} {arXiv:2507.23212
  [astro-ph.GA]} \BibitemShut {NoStop}%
\bibitem [{\citenamefont {Vijaykumar}\ \emph {et~al.}(2023)\citenamefont
  {Vijaykumar}, \citenamefont {Tiwari}, \citenamefont {Kapadia}, \citenamefont
  {Arun},\ and\ \citenamefont {Ajith}}]{Vijaykumar:2023tjg}%
  \BibitemOpen
  \bibfield  {author} {\bibinfo {author} {\bibfnamefont {Aditya}\ \bibnamefont
  {Vijaykumar}}, \bibinfo {author} {\bibfnamefont {Avinash}\ \bibnamefont
  {Tiwari}}, \bibinfo {author} {\bibfnamefont {Shasvath~J.}\ \bibnamefont
  {Kapadia}}, \bibinfo {author} {\bibfnamefont {K.~G.}\ \bibnamefont {Arun}}, \
  and\ \bibinfo {author} {\bibfnamefont {Parameswaran}\ \bibnamefont {Ajith}},\
  }\bibfield  {title} {\enquote {\bibinfo {title} {{Waltzing Binaries: Probing
  the Line-of-sight Acceleration of Merging Compact Objects with Gravitational
  Waves}},}\ }\href {\doibase 10.3847/1538-4357/acd77d} {\bibfield  {journal}
  {\bibinfo  {journal} {Astrophys. J.}\ }\textbf {\bibinfo {volume} {954}},\
  \bibinfo {pages} {105} (\bibinfo {year} {2023})},\ \Eprint
  {http://arxiv.org/abs/2302.09651} {arXiv:2302.09651 [astro-ph.HE]}
  \BibitemShut {NoStop}%
\bibitem [{\citenamefont {Tiwari}\ \emph {et~al.}(2023)\citenamefont {Tiwari},
  \citenamefont {Vijaykumar}, \citenamefont {Kapadia}, \citenamefont
  {Fragione},\ and\ \citenamefont {Chatterjee}}]{Tiwari:2023cpa}%
  \BibitemOpen
  \bibfield  {author} {\bibinfo {author} {\bibfnamefont {Avinash}\ \bibnamefont
  {Tiwari}}, \bibinfo {author} {\bibfnamefont {Aditya}\ \bibnamefont
  {Vijaykumar}}, \bibinfo {author} {\bibfnamefont {Shasvath~J.}\ \bibnamefont
  {Kapadia}}, \bibinfo {author} {\bibfnamefont {Giacomo}\ \bibnamefont
  {Fragione}}, \ and\ \bibinfo {author} {\bibfnamefont {Sourav}\ \bibnamefont
  {Chatterjee}},\ }\bibfield  {title} {\enquote {\bibinfo {title} {{Accelerated
  binary black holes in globular clusters: forecasts and detectability in the
  era of space-based gravitational-wave detectors}},}\ }\href {\doibase
  10.1093/mnras/stad3749} {\  (\bibinfo {year} {2023}),\
  10.1093/mnras/stad3749},\ \Eprint {http://arxiv.org/abs/2307.00930}
  {arXiv:2307.00930 [astro-ph.HE]} \BibitemShut {NoStop}%
\bibitem [{\citenamefont {Vajpeyi}\ \emph {et~al.}(2022)\citenamefont
  {Vajpeyi}, \citenamefont {Thrane}, \citenamefont {Smith}, \citenamefont
  {McKernan},\ and\ \citenamefont {Ford}}]{Vajpeyi:2021qsw}%
  \BibitemOpen
  \bibfield  {author} {\bibinfo {author} {\bibfnamefont {Avi}\ \bibnamefont
  {Vajpeyi}}, \bibinfo {author} {\bibfnamefont {Eric}\ \bibnamefont {Thrane}},
  \bibinfo {author} {\bibfnamefont {Rory}\ \bibnamefont {Smith}}, \bibinfo
  {author} {\bibfnamefont {Barry}\ \bibnamefont {McKernan}}, \ and\ \bibinfo
  {author} {\bibfnamefont {K.~E.~Saavik}\ \bibnamefont {Ford}},\ }\bibfield
  {title} {\enquote {\bibinfo {title} {{Measuring the Properties of Active
  Galactic Nuclei Disks with Gravitational Waves}},}\ }\href {\doibase
  10.3847/1538-4357/ac6180} {\bibfield  {journal} {\bibinfo  {journal}
  {Astrophys. J.}\ }\textbf {\bibinfo {volume} {931}},\ \bibinfo {pages} {82}
  (\bibinfo {year} {2022})},\ \Eprint {http://arxiv.org/abs/2111.03992}
  {arXiv:2111.03992 [gr-qc]} \BibitemShut {NoStop}%
\bibitem [{\citenamefont {Gond\'an}\ and\ \citenamefont
  {Kocsis}(2021)}]{Gondan:2020svr}%
  \BibitemOpen
  \bibfield  {author} {\bibinfo {author} {\bibfnamefont {L\'aszl\'o}\
  \bibnamefont {Gond\'an}}\ and\ \bibinfo {author} {\bibfnamefont {Bence}\
  \bibnamefont {Kocsis}},\ }\bibfield  {title} {\enquote {\bibinfo {title}
  {{High eccentricities and high masses characterize gravitational-wave
  captures in galactic nuclei as seen by Earth-based detectors}},}\ }\href
  {\doibase 10.1093/mnras/stab1722} {\bibfield  {journal} {\bibinfo  {journal}
  {Mon. Not. Roy. Astron. Soc.}\ }\textbf {\bibinfo {volume} {506}},\ \bibinfo
  {pages} {1665--1696} (\bibinfo {year} {2021})},\ \Eprint
  {http://arxiv.org/abs/2011.02507} {arXiv:2011.02507 [astro-ph.HE]}
  \BibitemShut {NoStop}%
\bibitem [{\citenamefont {Peters}\ and\ \citenamefont
  {Mathews}(1963)}]{Peters:1963ux}%
  \BibitemOpen
  \bibfield  {author} {\bibinfo {author} {\bibfnamefont {P.~C.}\ \bibnamefont
  {Peters}}\ and\ \bibinfo {author} {\bibfnamefont {J.}~\bibnamefont
  {Mathews}},\ }\bibfield  {title} {\enquote {\bibinfo {title} {{Gravitational
  radiation from point masses in a Keplerian orbit}},}\ }\href {\doibase
  10.1103/PhysRev.131.435} {\bibfield  {journal} {\bibinfo  {journal} {Phys.
  Rev.}\ }\textbf {\bibinfo {volume} {131}},\ \bibinfo {pages} {435--439}
  (\bibinfo {year} {1963})}\BibitemShut {NoStop}%
\bibitem [{\citenamefont {Samsing}\ \emph {et~al.}(2022)\citenamefont
  {Samsing}, \citenamefont {Bartos}, \citenamefont {D'Orazio}, \citenamefont
  {Haiman}, \citenamefont {Kocsis}, \citenamefont {Leigh}, \citenamefont {Liu},
  \citenamefont {Pessah},\ and\ \citenamefont {Tagawa}}]{Samsing:2020tda}%
  \BibitemOpen
  \bibfield  {author} {\bibinfo {author} {\bibfnamefont {J.}~\bibnamefont
  {Samsing}}, \bibinfo {author} {\bibfnamefont {I.}~\bibnamefont {Bartos}},
  \bibinfo {author} {\bibfnamefont {D.~J.}\ \bibnamefont {D'Orazio}}, \bibinfo
  {author} {\bibfnamefont {Z.}~\bibnamefont {Haiman}}, \bibinfo {author}
  {\bibfnamefont {B.}~\bibnamefont {Kocsis}}, \bibinfo {author} {\bibfnamefont
  {N.~W.~C.}\ \bibnamefont {Leigh}}, \bibinfo {author} {\bibfnamefont
  {B.}~\bibnamefont {Liu}}, \bibinfo {author} {\bibfnamefont {M.~E.}\
  \bibnamefont {Pessah}}, \ and\ \bibinfo {author} {\bibfnamefont
  {H.}~\bibnamefont {Tagawa}},\ }\bibfield  {title} {\enquote {\bibinfo {title}
  {{AGN as potential factories for eccentric black hole mergers}},}\ }\href
  {\doibase 10.1038/s41586-021-04333-1} {\bibfield  {journal} {\bibinfo
  {journal} {Nature}\ }\textbf {\bibinfo {volume} {603}},\ \bibinfo {pages}
  {237--240} (\bibinfo {year} {2022})},\ \Eprint
  {http://arxiv.org/abs/2010.09765} {arXiv:2010.09765 [astro-ph.HE]}
  \BibitemShut {NoStop}%
\bibitem [{\citenamefont {Tagawa}\ \emph {et~al.}(2021)\citenamefont {Tagawa},
  \citenamefont {Kocsis}, \citenamefont {Haiman}, \citenamefont {Bartos},
  \citenamefont {Omukai},\ and\ \citenamefont {Samsing}}]{tagawa2021eccentric}%
  \BibitemOpen
  \bibfield  {author} {\bibinfo {author} {\bibfnamefont {Hiromichi}\
  \bibnamefont {Tagawa}}, \bibinfo {author} {\bibfnamefont {Bence}\
  \bibnamefont {Kocsis}}, \bibinfo {author} {\bibfnamefont {Zolt{\'a}n}\
  \bibnamefont {Haiman}}, \bibinfo {author} {\bibfnamefont {Imre}\ \bibnamefont
  {Bartos}}, \bibinfo {author} {\bibfnamefont {Kazuyuki}\ \bibnamefont
  {Omukai}}, \ and\ \bibinfo {author} {\bibfnamefont {Johan}\ \bibnamefont
  {Samsing}},\ }\bibfield  {title} {\enquote {\bibinfo {title} {Eccentric black
  hole mergers in active galactic nuclei},}\ }\href@noop {} {\bibfield
  {journal} {\bibinfo  {journal} {The Astrophysical Journal Letters}\ }\textbf
  {\bibinfo {volume} {907}},\ \bibinfo {pages} {L20} (\bibinfo {year}
  {2021})}\BibitemShut {NoStop}%
\bibitem [{\citenamefont {Kozai}(1962)}]{Kozai:1962zz}%
  \BibitemOpen
  \bibfield  {author} {\bibinfo {author} {\bibfnamefont {Yoshihide}\
  \bibnamefont {Kozai}},\ }\bibfield  {title} {\enquote {\bibinfo {title}
  {{Secular perturbations of asteroids with high inclination and
  eccentricity}},}\ }\href {\doibase 10.1086/108790} {\bibfield  {journal}
  {\bibinfo  {journal} {Astron. J.}\ }\textbf {\bibinfo {volume} {67}},\
  \bibinfo {pages} {591--598} (\bibinfo {year} {1962})}\BibitemShut {NoStop}%
\bibitem [{\citenamefont {Lidov}(1962)}]{Lidov:1962wjn}%
  \BibitemOpen
  \bibfield  {author} {\bibinfo {author} {\bibfnamefont {M.~L.}\ \bibnamefont
  {Lidov}},\ }\bibfield  {title} {\enquote {\bibinfo {title} {{The evolution of
  orbits of artificial satellites of planets under the action of gravitational
  perturbations of external bodies}},}\ }\href {\doibase
  10.1016/0032-0633(62)90129-0} {\bibfield  {journal} {\bibinfo  {journal}
  {Planet. Space Sci.}\ }\textbf {\bibinfo {volume} {9}},\ \bibinfo {pages}
  {719--759} (\bibinfo {year} {1962})}\BibitemShut {NoStop}%
\bibitem [{\citenamefont {{Naoz}}(2016)}]{Naoz:2016tri}%
  \BibitemOpen
  \bibfield  {author} {\bibinfo {author} {\bibfnamefont {Smadar}\ \bibnamefont
  {{Naoz}}},\ }\bibfield  {title} {\enquote {\bibinfo {title} {{The Eccentric
  Kozai-Lidov Effect and Its Applications}},}\ }\href {\doibase
  10.1146/annurev-astro-081915-023315} {\bibfield  {journal} {\bibinfo
  {journal} {Annual Review of Astronomy and Astrophysics}\ }\textbf {\bibinfo
  {volume} {54}},\ \bibinfo {pages} {441--489} (\bibinfo {year} {2016})},\
  \Eprint {http://arxiv.org/abs/1601.07175} {arXiv:1601.07175 [astro-ph.EP]}
  \BibitemShut {NoStop}%
\bibitem [{\citenamefont {Antonini}\ \emph {et~al.}(2017)\citenamefont
  {Antonini}, \citenamefont {Toonen},\ and\ \citenamefont
  {Hamers}}]{Antonini:2017ash}%
  \BibitemOpen
  \bibfield  {author} {\bibinfo {author} {\bibfnamefont {Fabio}\ \bibnamefont
  {Antonini}}, \bibinfo {author} {\bibfnamefont {Silvia}\ \bibnamefont
  {Toonen}}, \ and\ \bibinfo {author} {\bibfnamefont {Adrian~S.}\ \bibnamefont
  {Hamers}},\ }\bibfield  {title} {\enquote {\bibinfo {title} {{Binary black
  hole mergers from field triples: properties, rates and the impact of stellar
  evolution}},}\ }\href {\doibase 10.3847/1538-4357/aa6f5e} {\bibfield
  {journal} {\bibinfo  {journal} {Astrophys. J.}\ }\textbf {\bibinfo {volume}
  {841}},\ \bibinfo {pages} {77} (\bibinfo {year} {2017})},\ \Eprint
  {http://arxiv.org/abs/1703.06614} {arXiv:1703.06614 [astro-ph.GA]}
  \BibitemShut {NoStop}%
\bibitem [{\citenamefont {Romero-Shaw}\ \emph {et~al.}(2019)\citenamefont
  {Romero-Shaw}, \citenamefont {Lasky},\ and\ \citenamefont
  {Thrane}}]{Romero-Shaw:2019itr}%
  \BibitemOpen
  \bibfield  {author} {\bibinfo {author} {\bibfnamefont {Isobel~M.}\
  \bibnamefont {Romero-Shaw}}, \bibinfo {author} {\bibfnamefont {Paul~D.}\
  \bibnamefont {Lasky}}, \ and\ \bibinfo {author} {\bibfnamefont {Eric}\
  \bibnamefont {Thrane}},\ }\bibfield  {title} {\enquote {\bibinfo {title}
  {{Searching for Eccentricity: Signatures of Dynamical Formation in the First
  Gravitational-Wave Transient Catalogue of LIGO and Virgo}},}\ }\href
  {\doibase 10.1093/mnras/stz2996} {\bibfield  {journal} {\bibinfo  {journal}
  {Mon. Not. Roy. Astron. Soc.}\ }\textbf {\bibinfo {volume} {490}},\ \bibinfo
  {pages} {5210--5216} (\bibinfo {year} {2019})},\ \Eprint
  {http://arxiv.org/abs/1909.05466} {arXiv:1909.05466 [astro-ph.HE]}
  \BibitemShut {NoStop}%
\bibitem [{\citenamefont {Romero-Shaw}\ \emph {et~al.}(2021)\citenamefont
  {Romero-Shaw}, \citenamefont {Lasky},\ and\ \citenamefont
  {Thrane}}]{Romero-Shaw:2021ual}%
  \BibitemOpen
  \bibfield  {author} {\bibinfo {author} {\bibfnamefont {Isobel~M.}\
  \bibnamefont {Romero-Shaw}}, \bibinfo {author} {\bibfnamefont {Paul~D.}\
  \bibnamefont {Lasky}}, \ and\ \bibinfo {author} {\bibfnamefont {Eric}\
  \bibnamefont {Thrane}},\ }\bibfield  {title} {\enquote {\bibinfo {title}
  {{Signs of Eccentricity in Two Gravitational-wave Signals May Indicate a
  Subpopulation of Dynamically Assembled Binary Black Holes}},}\ }\href
  {\doibase 10.3847/2041-8213/ac3138} {\bibfield  {journal} {\bibinfo
  {journal} {Astrophys. J. Lett.}\ }\textbf {\bibinfo {volume} {921}},\
  \bibinfo {pages} {L31} (\bibinfo {year} {2021})},\ \Eprint
  {http://arxiv.org/abs/2108.01284} {arXiv:2108.01284 [astro-ph.HE]}
  \BibitemShut {NoStop}%
\bibitem [{\citenamefont {Iglesias}\ \emph {et~al.}(2024)\citenamefont
  {Iglesias} \emph {et~al.}}]{Iglesias:2022xfc}%
  \BibitemOpen
  \bibfield  {author} {\bibinfo {author} {\bibfnamefont {H.~L.}\ \bibnamefont
  {Iglesias}} \emph {et~al.},\ }\bibfield  {title} {\enquote {\bibinfo {title}
  {{Eccentricity Estimation for Five Binary Black Hole Mergers with
  Higher-order Gravitational-wave Modes}},}\ }\href {\doibase
  10.3847/1538-4357/ad5ff6} {\bibfield  {journal} {\bibinfo  {journal}
  {Astrophys. J.}\ }\textbf {\bibinfo {volume} {972}},\ \bibinfo {pages} {65}
  (\bibinfo {year} {2024})},\ \Eprint {http://arxiv.org/abs/2208.01766}
  {arXiv:2208.01766 [gr-qc]} \BibitemShut {NoStop}%
\bibitem [{\citenamefont {Wu}\ \emph {et~al.}(2020)\citenamefont {Wu},
  \citenamefont {Cao},\ and\ \citenamefont {Zhu}}]{Wu:2020zwr}%
  \BibitemOpen
  \bibfield  {author} {\bibinfo {author} {\bibfnamefont {Shichao}\ \bibnamefont
  {Wu}}, \bibinfo {author} {\bibfnamefont {Zhoujian}\ \bibnamefont {Cao}}, \
  and\ \bibinfo {author} {\bibfnamefont {Zong-Hong}\ \bibnamefont {Zhu}},\
  }\bibfield  {title} {\enquote {\bibinfo {title} {{Measuring the eccentricity
  of binary black holes in GWTC-1 by using the inspiral-only waveform}},}\
  }\href {\doibase 10.1093/mnras/staa1176} {\bibfield  {journal} {\bibinfo
  {journal} {Mon. Not. Roy. Astron. Soc.}\ }\textbf {\bibinfo {volume} {495}},\
  \bibinfo {pages} {466--478} (\bibinfo {year} {2020})},\ \Eprint
  {http://arxiv.org/abs/2002.05528} {arXiv:2002.05528 [astro-ph.IM]}
  \BibitemShut {NoStop}%
\bibitem [{\citenamefont {Planas}\ \emph {et~al.}(2025)\citenamefont {Planas},
  \citenamefont {Ramos-Buades}, \citenamefont {Garc{\'\i}a-Quir{\'o}s},
  \citenamefont {Estell{\'e}s}, \citenamefont {Husa},\ and\ \citenamefont
  {Haney}}]{Planas:2025jny}%
  \BibitemOpen
  \bibfield  {author} {\bibinfo {author} {\bibfnamefont {Maria de~Lluc}\
  \bibnamefont {Planas}}, \bibinfo {author} {\bibfnamefont {Antoni}\
  \bibnamefont {Ramos-Buades}}, \bibinfo {author} {\bibfnamefont {Cecilio}\
  \bibnamefont {Garc{\'\i}a-Quir{\'o}s}}, \bibinfo {author} {\bibfnamefont
  {H{\'e}ctor}\ \bibnamefont {Estell{\'e}s}}, \bibinfo {author} {\bibfnamefont
  {Sascha}\ \bibnamefont {Husa}}, \ and\ \bibinfo {author} {\bibfnamefont
  {Maria}\ \bibnamefont {Haney}},\ }\bibfield  {title} {\enquote {\bibinfo
  {title} {{Eccentric or circular? A reanalysis of binary black hole
  gravitational wave events for orbital eccentricity signatures}},}\
  }\href@noop {} {\  (\bibinfo {year} {2025})},\ \Eprint
  {http://arxiv.org/abs/2504.15833} {arXiv:2504.15833 [gr-qc]} \BibitemShut
  {NoStop}%
\bibitem [{\citenamefont {Morras}\ \emph {et~al.}(2025)\citenamefont {Morras},
  \citenamefont {Pratten},\ and\ \citenamefont
  {Schmidt}}]{Morras2025_GW200105_ecc}%
  \BibitemOpen
  \bibfield  {author} {\bibinfo {author} {\bibfnamefont {Gonzalo}\ \bibnamefont
  {Morras}}, \bibinfo {author} {\bibfnamefont {Geraint}\ \bibnamefont
  {Pratten}}, \ and\ \bibinfo {author} {\bibfnamefont {Patricia}\ \bibnamefont
  {Schmidt}},\ }\bibfield  {title} {\enquote {\bibinfo {title} {{Orbital
  eccentricity in a neutron star -- black hole binary}},}\ }\href {\doibase
  10.48550/arXiv.2503.15393} {\bibfield  {journal} {\bibinfo  {journal} {arXiv
  preprint arXiv:2503.15393}\ } (\bibinfo {year} {2025}),\
  10.48550/arXiv.2503.15393}\BibitemShut {NoStop}%
\bibitem [{\citenamefont {Shaikh}\ \emph {et~al.}(2024)\citenamefont {Shaikh},
  \citenamefont {Bhat},\ and\ \citenamefont {Kapadia}}]{Shaikh:2024wyn}%
  \BibitemOpen
  \bibfield  {author} {\bibinfo {author} {\bibfnamefont {Md~Arif}\ \bibnamefont
  {Shaikh}}, \bibinfo {author} {\bibfnamefont {Sajad~A.}\ \bibnamefont {Bhat}},
  \ and\ \bibinfo {author} {\bibfnamefont {Shasvath~J.}\ \bibnamefont
  {Kapadia}},\ }\bibfield  {title} {\enquote {\bibinfo {title} {{A study of the
  inspiral-merger-ringdown consistency test with gravitational-wave signals
  from compact binaries in eccentric orbits}},}\ }\href {\doibase
  10.1103/PhysRevD.110.024030} {\bibfield  {journal} {\bibinfo  {journal}
  {Phys. Rev. D}\ }\textbf {\bibinfo {volume} {110}},\ \bibinfo {pages}
  {024030} (\bibinfo {year} {2024})},\ \Eprint
  {http://arxiv.org/abs/2402.15110} {arXiv:2402.15110 [gr-qc]} \BibitemShut
  {NoStop}%
\bibitem [{\citenamefont {Bhat}\ \emph {et~al.}(2023)\citenamefont {Bhat},
  \citenamefont {Saini}, \citenamefont {Favata},\ and\ \citenamefont
  {Arun}}]{Bhat:2022amc}%
  \BibitemOpen
  \bibfield  {author} {\bibinfo {author} {\bibfnamefont {Sajad~A.}\
  \bibnamefont {Bhat}}, \bibinfo {author} {\bibfnamefont {Pankaj}\ \bibnamefont
  {Saini}}, \bibinfo {author} {\bibfnamefont {Marc}\ \bibnamefont {Favata}}, \
  and\ \bibinfo {author} {\bibfnamefont {K.~G.}\ \bibnamefont {Arun}},\
  }\bibfield  {title} {\enquote {\bibinfo {title} {{Systematic bias on the
  inspiral-merger-ringdown consistency test due to neglect of orbital
  eccentricity}},}\ }\href {\doibase 10.1103/PhysRevD.107.024009} {\bibfield
  {journal} {\bibinfo  {journal} {Phys. Rev. D}\ }\textbf {\bibinfo {volume}
  {107}},\ \bibinfo {pages} {024009} (\bibinfo {year} {2023})},\ \Eprint
  {http://arxiv.org/abs/2207.13761} {arXiv:2207.13761 [gr-qc]} \BibitemShut
  {NoStop}%
\bibitem [{\citenamefont {Saini}\ \emph {et~al.}(2022)\citenamefont {Saini},
  \citenamefont {Favata},\ and\ \citenamefont {Arun}}]{Saini:2022igm}%
  \BibitemOpen
  \bibfield  {author} {\bibinfo {author} {\bibfnamefont {Pankaj}\ \bibnamefont
  {Saini}}, \bibinfo {author} {\bibfnamefont {Marc}\ \bibnamefont {Favata}}, \
  and\ \bibinfo {author} {\bibfnamefont {K.~G.}\ \bibnamefont {Arun}},\
  }\bibfield  {title} {\enquote {\bibinfo {title} {{Systematic bias on
  parametrized tests of general relativity due to neglect of orbital
  eccentricity}},}\ }\href {\doibase 10.1103/PhysRevD.106.084031} {\bibfield
  {journal} {\bibinfo  {journal} {Phys. Rev. D}\ }\textbf {\bibinfo {volume}
  {106}},\ \bibinfo {pages} {084031} (\bibinfo {year} {2022})},\ \Eprint
  {http://arxiv.org/abs/2203.04634} {arXiv:2203.04634 [gr-qc]} \BibitemShut
  {NoStop}%
\bibitem [{\citenamefont {Saini}\ \emph {et~al.}(2024)\citenamefont {Saini},
  \citenamefont {Bhat}, \citenamefont {Favata},\ and\ \citenamefont
  {Arun}}]{Saini:2023rto}%
  \BibitemOpen
  \bibfield  {author} {\bibinfo {author} {\bibfnamefont {Pankaj}\ \bibnamefont
  {Saini}}, \bibinfo {author} {\bibfnamefont {Sajad~A.}\ \bibnamefont {Bhat}},
  \bibinfo {author} {\bibfnamefont {Marc}\ \bibnamefont {Favata}}, \ and\
  \bibinfo {author} {\bibfnamefont {K.~G.}\ \bibnamefont {Arun}},\ }\bibfield
  {title} {\enquote {\bibinfo {title} {{Eccentricity-induced systematic error
  on parametrized tests of general relativity: Hierarchical Bayesian inference
  applied to a binary black hole population}},}\ }\href {\doibase
  10.1103/PhysRevD.109.084056} {\bibfield  {journal} {\bibinfo  {journal}
  {Phys. Rev. D}\ }\textbf {\bibinfo {volume} {109}},\ \bibinfo {pages}
  {084056} (\bibinfo {year} {2024})},\ \Eprint
  {http://arxiv.org/abs/2311.08033} {arXiv:2311.08033 [gr-qc]} \BibitemShut
  {NoStop}%
\bibitem [{\citenamefont {Narayan}\ \emph {et~al.}(2023)\citenamefont
  {Narayan}, \citenamefont {Johnson-McDaniel},\ and\ \citenamefont
  {Gupta}}]{Narayan:2023vhm}%
  \BibitemOpen
  \bibfield  {author} {\bibinfo {author} {\bibfnamefont {Purnima}\ \bibnamefont
  {Narayan}}, \bibinfo {author} {\bibfnamefont {Nathan~K.}\ \bibnamefont
  {Johnson-McDaniel}}, \ and\ \bibinfo {author} {\bibfnamefont {Anuradha}\
  \bibnamefont {Gupta}},\ }\bibfield  {title} {\enquote {\bibinfo {title}
  {{Effect of ignoring eccentricity in testing general relativity with
  gravitational waves}},}\ }\href {\doibase 10.1103/PhysRevD.108.064003}
  {\bibfield  {journal} {\bibinfo  {journal} {Phys. Rev. D}\ }\textbf {\bibinfo
  {volume} {108}},\ \bibinfo {pages} {064003} (\bibinfo {year} {2023})},\
  \Eprint {http://arxiv.org/abs/2306.04068} {arXiv:2306.04068 [gr-qc]}
  \BibitemShut {NoStop}%
\bibitem [{\citenamefont {Tiwari}\ \emph {et~al.}(2025)\citenamefont {Tiwari},
  \citenamefont {Vijaykumar}, \citenamefont {Kapadia}, \citenamefont {Ghosh},\
  and\ \citenamefont {Nielsen}}]{Tiwari:2025aec}%
  \BibitemOpen
  \bibfield  {author} {\bibinfo {author} {\bibfnamefont {Avinash}\ \bibnamefont
  {Tiwari}}, \bibinfo {author} {\bibfnamefont {Aditya}\ \bibnamefont
  {Vijaykumar}}, \bibinfo {author} {\bibfnamefont {Shasvath~J.}\ \bibnamefont
  {Kapadia}}, \bibinfo {author} {\bibfnamefont {Shrobana}\ \bibnamefont
  {Ghosh}}, \ and\ \bibinfo {author} {\bibfnamefont {Alex~B.}\ \bibnamefont
  {Nielsen}},\ }\bibfield  {title} {\enquote {\bibinfo {title} {{A pipeline to
  search for signatures of line-of-sight acceleration in gravitational wave
  signals produced by compact binary coalescences}},}\ }\href@noop {} {\
  (\bibinfo {year} {2025})},\ \Eprint {http://arxiv.org/abs/2506.22272}
  {arXiv:2506.22272 [astro-ph.HE]} \BibitemShut {NoStop}%
\bibitem [{\citenamefont {Romero-Shaw}\ \emph
  {et~al.}(2023{\natexlab{a}})\citenamefont {Romero-Shaw}, \citenamefont
  {Gerosa},\ and\ \citenamefont
  {Loutrel}}]{RomeroShaw2023_precession_eccentricity_degeneracy}%
  \BibitemOpen
  \bibfield  {author} {\bibinfo {author} {\bibfnamefont {Isobel~M.}\
  \bibnamefont {Romero-Shaw}}, \bibinfo {author} {\bibfnamefont {Davide}\
  \bibnamefont {Gerosa}}, \ and\ \bibinfo {author} {\bibfnamefont {Nicholas}\
  \bibnamefont {Loutrel}},\ }\bibfield  {title} {\enquote {\bibinfo {title}
  {{Eccentricity or spin precession? Distinguishing subdominant effects in
  gravitational-wave data}},}\ }\href {\doibase 10.1093/mnras/stad031}
  {\bibfield  {journal} {\bibinfo  {journal} {Monthly Notices of the Royal
  Astronomical Society}\ }\textbf {\bibinfo {volume} {519}},\ \bibinfo {pages}
  {5352--5357} (\bibinfo {year} {2023}{\natexlab{a}})},\ \Eprint
  {http://arxiv.org/abs/2211.07528} {arXiv:2211.07528 [astro-ph.HE]}
  \BibitemShut {NoStop}%
\bibitem [{\citenamefont {et~al.}(2025)}]{mishra-ecc-mclz}%
  \BibitemOpen
  \bibfield  {author} {\bibinfo {author} {\bibfnamefont {Anuj~Mishra}\
  \bibnamefont {et~al.}},\ }\href@noop {} {\enquote {\bibinfo {title}
  {Degeneracy between eccentricity and microlensing effects in compact binary
  gravitational signals},}\ }\bibinfo {howpublished} {Unpublished work}
  (\bibinfo {year} {2025}),\ \bibinfo {note} {in preparation}\BibitemShut
  {NoStop}%
\bibitem [{\citenamefont {Clarke}\ \emph {et~al.}(2022)\citenamefont {Clarke},
  \citenamefont {Romero-Shaw}, \citenamefont {Lasky},\ and\ \citenamefont
  {Thrane}}]{Clarke:2022fma}%
  \BibitemOpen
  \bibfield  {author} {\bibinfo {author} {\bibfnamefont {Teagan~A.}\
  \bibnamefont {Clarke}}, \bibinfo {author} {\bibfnamefont {Isobel~M.}\
  \bibnamefont {Romero-Shaw}}, \bibinfo {author} {\bibfnamefont {Paul~D.}\
  \bibnamefont {Lasky}}, \ and\ \bibinfo {author} {\bibfnamefont {Eric}\
  \bibnamefont {Thrane}},\ }\bibfield  {title} {\enquote {\bibinfo {title}
  {{Gravitational-wave inference for eccentric binaries: the argument of
  periapsis}},}\ }\href {\doibase 10.1093/mnras/stac2965} {\bibfield  {journal}
  {\bibinfo  {journal} {Mon. Not. Roy. Astron. Soc.}\ }\textbf {\bibinfo
  {volume} {517}},\ \bibinfo {pages} {3778--3784} (\bibinfo {year} {2022})},\
  \Eprint {http://arxiv.org/abs/2206.14006} {arXiv:2206.14006 [gr-qc]}
  \BibitemShut {NoStop}%
\bibitem [{\citenamefont {Ashton}\ \emph {et~al.}(2019)\citenamefont {Ashton},
  \citenamefont {H{\"u}bner}, \citenamefont {Lasky}, \citenamefont {Talbot},
  \citenamefont {Thrane} \emph {et~al.}}]{Ashton:2018jfp}%
  \BibitemOpen
  \bibfield  {author} {\bibinfo {author} {\bibfnamefont {Greg}\ \bibnamefont
  {Ashton}}, \bibinfo {author} {\bibfnamefont {Moritz}\ \bibnamefont
  {H{\"u}bner}}, \bibinfo {author} {\bibfnamefont {Paul~D.}\ \bibnamefont
  {Lasky}}, \bibinfo {author} {\bibfnamefont {Colin}\ \bibnamefont {Talbot}},
  \bibinfo {author} {\bibfnamefont {Eric}\ \bibnamefont {Thrane}},  \emph
  {et~al.},\ }\bibfield  {title} {\enquote {\bibinfo {title} {{Bilby: A
  User-friendly Bayesian Inference Library for Gravitational-wave
  Astronomy}},}\ }\href {\doibase 10.3847/1538-4365/ab06fc} {\bibfield
  {journal} {\bibinfo  {journal} {Astrophys. J. Suppl.}\ }\textbf {\bibinfo
  {volume} {241}},\ \bibinfo {pages} {27} (\bibinfo {year} {2019})},\ \Eprint
  {http://arxiv.org/abs/1811.02042} {arXiv:1811.02042 [astro-ph.IM]}
  \BibitemShut {NoStop}%
\bibitem [{\citenamefont {Smith}\ \emph {et~al.}(2020)\citenamefont {Smith},
  \citenamefont {Ashton}, \citenamefont {Vajpeyi},\ and\ \citenamefont
  {Lasky}}]{Smith:2020}%
  \BibitemOpen
  \bibfield  {author} {\bibinfo {author} {\bibfnamefont {Rory}\ \bibnamefont
  {Smith}}, \bibinfo {author} {\bibfnamefont {Greg}\ \bibnamefont {Ashton}},
  \bibinfo {author} {\bibfnamefont {Aditya}\ \bibnamefont {Vajpeyi}}, \ and\
  \bibinfo {author} {\bibfnamefont {Paul~D.}\ \bibnamefont {Lasky}},\
  }\bibfield  {title} {\enquote {\bibinfo {title} {{bilby\_pipe: A
  user-friendly tool for distributed parameter estimation using bilby}},}\
  }\href {\doibase 10.21105/joss.02895} {\bibfield  {journal} {\bibinfo
  {journal} {Journal of Open Source Software}\ }\textbf {\bibinfo {volume}
  {5}},\ \bibinfo {pages} {2895} (\bibinfo {year} {2020})}\BibitemShut
  {NoStop}%
\bibitem [{\citenamefont {Speagle}(2020)}]{Speagle:2019ivv}%
  \BibitemOpen
  \bibfield  {author} {\bibinfo {author} {\bibfnamefont {Joshua~S.}\
  \bibnamefont {Speagle}},\ }\bibfield  {title} {\enquote {\bibinfo {title}
  {{dynesty: a dynamic nested sampling package for estimating Bayesian
  posteriors and evidences}},}\ }\href {\doibase 10.1093/mnras/staa278}
  {\bibfield  {journal} {\bibinfo  {journal} {Mon. Not. Roy. Astron. Soc.}\
  }\textbf {\bibinfo {volume} {493}},\ \bibinfo {pages} {3132--3158} (\bibinfo
  {year} {2020})},\ \Eprint {http://arxiv.org/abs/1904.02180} {arXiv:1904.02180
  [astro-ph.IM]} \BibitemShut {NoStop}%
\bibitem [{\citenamefont {Buonanno}\ \emph {et~al.}(2009)\citenamefont
  {Buonanno}, \citenamefont {Iyer}, \citenamefont {Ochsner}, \citenamefont
  {Pan},\ and\ \citenamefont {Sathyaprakash}}]{Buonanno:2009zt}%
  \BibitemOpen
  \bibfield  {author} {\bibinfo {author} {\bibfnamefont {Alessandra}\
  \bibnamefont {Buonanno}}, \bibinfo {author} {\bibfnamefont {Bala}\
  \bibnamefont {Iyer}}, \bibinfo {author} {\bibfnamefont {Evan}\ \bibnamefont
  {Ochsner}}, \bibinfo {author} {\bibfnamefont {Yi}~\bibnamefont {Pan}}, \ and\
  \bibinfo {author} {\bibfnamefont {B.~S.}\ \bibnamefont {Sathyaprakash}},\
  }\bibfield  {title} {\enquote {\bibinfo {title} {{Comparison of
  post-Newtonian templates for compact binary inspiral signals in
  gravitational-wave detectors}},}\ }\href {\doibase
  10.1103/PhysRevD.80.084043} {\bibfield  {journal} {\bibinfo  {journal} {Phys.
  Rev. D}\ }\textbf {\bibinfo {volume} {80}},\ \bibinfo {pages} {084043}
  (\bibinfo {year} {2009})},\ \Eprint {http://arxiv.org/abs/0907.0700}
  {arXiv:0907.0700 [gr-qc]} \BibitemShut {NoStop}%
\bibitem [{\citenamefont {Takahashi}\ and\ \citenamefont
  {Nakamura}(2003)}]{Takahashi:2003ay}%
  \BibitemOpen
  \bibfield  {author} {\bibinfo {author} {\bibfnamefont {R.}~\bibnamefont
  {Takahashi}}\ and\ \bibinfo {author} {\bibfnamefont {T.}~\bibnamefont
  {Nakamura}},\ }\bibfield  {title} {\enquote {\bibinfo {title} {{Wave effects
  in gravitational lensing of gravitational waves from chirping binaries}},}\
  }in\ \href@noop {} {\emph {\bibinfo {booktitle} {{28th International Cosmic
  Ray Conference}}}}\ (\bibinfo {year} {2003})\ pp.\ \bibinfo {pages}
  {3153--3156}\BibitemShut {NoStop}%
\bibitem [{\citenamefont {Dai}\ and\ \citenamefont
  {Venumadhav}(2017)}]{Dai:2017huk}%
  \BibitemOpen
  \bibfield  {author} {\bibinfo {author} {\bibfnamefont {Liang}\ \bibnamefont
  {Dai}}\ and\ \bibinfo {author} {\bibfnamefont {Tejaswi}\ \bibnamefont
  {Venumadhav}},\ }\bibfield  {title} {\enquote {\bibinfo {title} {{On the
  waveforms of gravitationally lensed gravitational waves}},}\ }\href@noop {}
  {\  (\bibinfo {year} {2017})},\ \Eprint {http://arxiv.org/abs/1702.04724}
  {arXiv:1702.04724 [gr-qc]} \BibitemShut {NoStop}%
\bibitem [{\citenamefont {Yeung}\ \emph {et~al.}(2023)\citenamefont {Yeung},
  \citenamefont {Cheung}, \citenamefont {Seo}, \citenamefont {Gais},
  \citenamefont {Hannuksela},\ and\ \citenamefont {Li}}]{Yeung:2021chy}%
  \BibitemOpen
  \bibfield  {author} {\bibinfo {author} {\bibfnamefont {Simon M.~C.}\
  \bibnamefont {Yeung}}, \bibinfo {author} {\bibfnamefont {Mark H.~Y.}\
  \bibnamefont {Cheung}}, \bibinfo {author} {\bibfnamefont {Eungwang}\
  \bibnamefont {Seo}}, \bibinfo {author} {\bibfnamefont {Joseph A.~J.}\
  \bibnamefont {Gais}}, \bibinfo {author} {\bibfnamefont {Otto~A.}\
  \bibnamefont {Hannuksela}}, \ and\ \bibinfo {author} {\bibfnamefont {Tjonnie
  G.~F.}\ \bibnamefont {Li}},\ }\bibfield  {title} {\enquote {\bibinfo {title}
  {{Detectability of microlensed gravitational waves}},}\ }\href {\doibase
  10.1093/mnras/stad2772} {\bibfield  {journal} {\bibinfo  {journal} {Mon. Not.
  Roy. Astron. Soc.}\ }\textbf {\bibinfo {volume} {526}},\ \bibinfo {pages}
  {2230--2240} (\bibinfo {year} {2023})},\ \Eprint
  {http://arxiv.org/abs/2112.07635} {arXiv:2112.07635 [gr-qc]} \BibitemShut
  {NoStop}%
\bibitem [{\citenamefont {Mishra}\ \emph {et~al.}(2021)\citenamefont {Mishra},
  \citenamefont {Meena}, \citenamefont {More}, \citenamefont {Bose},\ and\
  \citenamefont {Bagla}}]{Mishra:2021xzz}%
  \BibitemOpen
  \bibfield  {author} {\bibinfo {author} {\bibfnamefont {Anuj}\ \bibnamefont
  {Mishra}}, \bibinfo {author} {\bibfnamefont {Ashish~Kumar}\ \bibnamefont
  {Meena}}, \bibinfo {author} {\bibfnamefont {Anupreeta}\ \bibnamefont {More}},
  \bibinfo {author} {\bibfnamefont {Sukanta}\ \bibnamefont {Bose}}, \ and\
  \bibinfo {author} {\bibfnamefont {Jasjeet~Singh}\ \bibnamefont {Bagla}},\
  }\bibfield  {title} {\enquote {\bibinfo {title} {{Gravitational lensing of
  gravitational waves: effect of microlens population in lensing galaxies}},}\
  }\href {\doibase 10.1093/mnras/stab2875} {\bibfield  {journal} {\bibinfo
  {journal} {Mon. Not. Roy. Astron. Soc.}\ }\textbf {\bibinfo {volume} {508}},\
  \bibinfo {pages} {4869--4886} (\bibinfo {year} {2021})},\ \Eprint
  {http://arxiv.org/abs/2102.03946} {arXiv:2102.03946 [astro-ph.CO]}
  \BibitemShut {NoStop}%
\bibitem [{\citenamefont {Bonvin}\ \emph {et~al.}(2017)\citenamefont {Bonvin},
  \citenamefont {Caprini},\ and\ \citenamefont {Sturani}}]{Bonvin:2016qxr}%
  \BibitemOpen
  \bibfield  {author} {\bibinfo {author} {\bibfnamefont {Camille}\ \bibnamefont
  {Bonvin}}, \bibinfo {author} {\bibfnamefont {Chiara}\ \bibnamefont
  {Caprini}}, \ and\ \bibinfo {author} {\bibfnamefont {Riccardo}\ \bibnamefont
  {Sturani}},\ }\bibfield  {title} {\enquote {\bibinfo {title} {{Effect of
  matter structure on the gravitational waveforms}},}\ }\href {\doibase
  10.1103/PhysRevD.95.024045} {\bibfield  {journal} {\bibinfo  {journal} {Phys.
  Rev. D}\ }\textbf {\bibinfo {volume} {95}},\ \bibinfo {pages} {024045}
  (\bibinfo {year} {2017})},\ \Eprint {http://arxiv.org/abs/1609.08093}
  {arXiv:1609.08093 [gr-qc]} \BibitemShut {NoStop}%
\bibitem [{\citenamefont {Inayoshi}\ \emph {et~al.}(2018)\citenamefont
  {Inayoshi}, \citenamefont {Kashiyama}, \citenamefont {Visbal},\ and\
  \citenamefont {Haiman}}]{Inayoshi:2017hgw}%
  \BibitemOpen
  \bibfield  {author} {\bibinfo {author} {\bibfnamefont {Kohei}\ \bibnamefont
  {Inayoshi}}, \bibinfo {author} {\bibfnamefont {Kazumi}\ \bibnamefont
  {Kashiyama}}, \bibinfo {author} {\bibfnamefont {Eli}\ \bibnamefont {Visbal}},
  \ and\ \bibinfo {author} {\bibfnamefont {Zoltan}\ \bibnamefont {Haiman}},\
  }\bibfield  {title} {\enquote {\bibinfo {title} {{Gravitational Wave
  Background from Black Hole Binaries in the Hierarchical Triple Systems}},}\
  }\href {\doibase 10.1093/mnras/sty1521} {\bibfield  {journal} {\bibinfo
  {journal} {MNRAS}\ }\textbf {\bibinfo {volume} {479}},\ \bibinfo {pages}
  {1180--1190} (\bibinfo {year} {2018})},\ \Eprint
  {http://arxiv.org/abs/1701.04823} {arXiv:1701.04823 [astro-ph.HE]}
  \BibitemShut {NoStop}%
\bibitem [{\citenamefont {Moore}\ \emph {et~al.}(2016)\citenamefont {Moore},
  \citenamefont {Favata}, \citenamefont {Arun},\ and\ \citenamefont
  {Mishra}}]{Moore:2016qxz}%
  \BibitemOpen
  \bibfield  {author} {\bibinfo {author} {\bibfnamefont {Blake}\ \bibnamefont
  {Moore}}, \bibinfo {author} {\bibfnamefont {Marc}\ \bibnamefont {Favata}},
  \bibinfo {author} {\bibfnamefont {K.~G.}\ \bibnamefont {Arun}}, \ and\
  \bibinfo {author} {\bibfnamefont {Chandra~Kant}\ \bibnamefont {Mishra}},\
  }\bibfield  {title} {\enquote {\bibinfo {title} {{Gravitational-wave phasing
  for low-eccentricity inspiralling compact binaries to 3PN order}},}\ }\href
  {\doibase 10.1103/PhysRevD.93.124061} {\bibfield  {journal} {\bibinfo
  {journal} {Phys. Rev. D}\ }\textbf {\bibinfo {volume} {93}},\ \bibinfo
  {pages} {124061} (\bibinfo {year} {2016})},\ \Eprint
  {http://arxiv.org/abs/1605.00304} {arXiv:1605.00304 [gr-qc]} \BibitemShut
  {NoStop}%
\bibitem [{\citenamefont {Randall}\ and\ \citenamefont
  {Xianyu}(2018)}]{Randall:2018nud}%
  \BibitemOpen
  \bibfield  {author} {\bibinfo {author} {\bibfnamefont {Lisa}\ \bibnamefont
  {Randall}}\ and\ \bibinfo {author} {\bibfnamefont {Zhong-Zhi}\ \bibnamefont
  {Xianyu}},\ }\bibfield  {title} {\enquote {\bibinfo {title} {{A direct probe
  of mass density near inspiraling binary black holes}},}\ }\href {\doibase
  10.3847/1538-4357/aad7f1} {\bibfield  {journal} {\bibinfo  {journal}
  {Astrophys. J.}\ }\textbf {\bibinfo {volume} {864}},\ \bibinfo {pages} {134}
  (\bibinfo {year} {2018})},\ \Eprint {http://arxiv.org/abs/1806.07898}
  {arXiv:1806.07898 [astro-ph.HE]} \BibitemShut {NoStop}%
\bibitem [{\citenamefont {Abbott}\ \emph
  {et~al.}(2019{\natexlab{b}})\citenamefont {Abbott}, \citenamefont {Abbott},
  \citenamefont {Abbott}, \citenamefont {Abraham} \emph
  {et~al.}}]{PhysRevD.100.104036}%
  \BibitemOpen
  \bibfield  {author} {\bibinfo {author} {\bibfnamefont {B.~P.}\ \bibnamefont
  {Abbott}}, \bibinfo {author} {\bibfnamefont {R.}~\bibnamefont {Abbott}},
  \bibinfo {author} {\bibfnamefont {T.~D.}\ \bibnamefont {Abbott}}, \bibinfo
  {author} {\bibfnamefont {S.}~\bibnamefont {Abraham}},  \emph {et~al.}
  (\bibinfo {collaboration} {The LIGO Scientific Collaboration and the Virgo
  Collaboration}),\ }\bibfield  {title} {\enquote {\bibinfo {title} {Tests of
  general relativity with the binary black hole signals from the ligo-virgo
  catalog gwtc-1},}\ }\href {\doibase 10.1103/PhysRevD.100.104036} {\bibfield
  {journal} {\bibinfo  {journal} {Phys. Rev. D}\ }\textbf {\bibinfo {volume}
  {100}},\ \bibinfo {pages} {104036} (\bibinfo {year}
  {2019}{\natexlab{b}})}\BibitemShut {NoStop}%
\bibitem [{\citenamefont {Will}(1998)}]{Will:1997bb}%
  \BibitemOpen
  \bibfield  {author} {\bibinfo {author} {\bibfnamefont {C.~M.}\ \bibnamefont
  {Will}},\ }\bibfield  {title} {\enquote {\bibinfo {title} {Bounding the mass
  of the graviton using gravitational-wave observations of inspiralling compact
  binaries},}\ }\href {\doibase 10.1103/PhysRevD.57.2061} {\bibfield  {journal}
  {\bibinfo  {journal} {Phys. Rev. D}\ }\textbf {\bibinfo {volume} {57}},\
  \bibinfo {pages} {2061--2068} (\bibinfo {year} {1998})},\ \Eprint
  {http://arxiv.org/abs/gr-qc/9709011} {arXiv:gr-qc/9709011} \BibitemShut
  {NoStop}%
\bibitem [{\citenamefont {Yunes}\ \emph {et~al.}(2011)\citenamefont {Yunes},
  \citenamefont {Kocsis}, \citenamefont {Loeb},\ and\ \citenamefont
  {Haiman}}]{Yunes:2011ws}%
  \BibitemOpen
  \bibfield  {author} {\bibinfo {author} {\bibfnamefont {Nicolas}\ \bibnamefont
  {Yunes}}, \bibinfo {author} {\bibfnamefont {Bence}\ \bibnamefont {Kocsis}},
  \bibinfo {author} {\bibfnamefont {Abraham}\ \bibnamefont {Loeb}}, \ and\
  \bibinfo {author} {\bibfnamefont {Zoltan}\ \bibnamefont {Haiman}},\
  }\bibfield  {title} {\enquote {\bibinfo {title} {{Imprint of Accretion
  Disk-Induced Migration on Gravitational Waves from Extreme Mass Ratio
  Inspirals}},}\ }\href {\doibase 10.1103/PhysRevLett.107.171103} {\bibfield
  {journal} {\bibinfo  {journal} {Phys. Rev. Lett.}\ }\textbf {\bibinfo
  {volume} {107}},\ \bibinfo {pages} {171103} (\bibinfo {year} {2011})},\
  \Eprint {http://arxiv.org/abs/1103.4609} {arXiv:1103.4609 [astro-ph.CO]}
  \BibitemShut {NoStop}%
\bibitem [{\citenamefont {Will}(2014)}]{Will:2014kxa}%
  \BibitemOpen
  \bibfield  {author} {\bibinfo {author} {\bibfnamefont {Clifford~M.}\
  \bibnamefont {Will}},\ }\bibfield  {title} {\enquote {\bibinfo {title} {{The
  Confrontation between General Relativity and Experiment}},}\ }\href {\doibase
  10.12942/lrr-2014-4} {\bibfield  {journal} {\bibinfo  {journal} {Living Rev.
  Rel.}\ }\textbf {\bibinfo {volume} {17}},\ \bibinfo {pages} {4} (\bibinfo
  {year} {2014})},\ \Eprint {http://arxiv.org/abs/1403.7377} {arXiv:1403.7377
  [gr-qc]} \BibitemShut {NoStop}%
\bibitem [{\citenamefont {Barausse}\ \emph {et~al.}(2016)\citenamefont
  {Barausse}, \citenamefont {Yunes},\ and\ \citenamefont
  {Chamberlain}}]{PhysRevLett.116.241104}%
  \BibitemOpen
  \bibfield  {author} {\bibinfo {author} {\bibfnamefont {Enrico}\ \bibnamefont
  {Barausse}}, \bibinfo {author} {\bibfnamefont {Nicol\'as}\ \bibnamefont
  {Yunes}}, \ and\ \bibinfo {author} {\bibfnamefont {Katie}\ \bibnamefont
  {Chamberlain}},\ }\bibfield  {title} {\enquote {\bibinfo {title}
  {Theory-agnostic constraints on black-hole dipole radiation with multiband
  gravitational-wave astrophysics},}\ }\href {\doibase
  10.1103/PhysRevLett.116.241104} {\bibfield  {journal} {\bibinfo  {journal}
  {Phys. Rev. Lett.}\ }\textbf {\bibinfo {volume} {116}},\ \bibinfo {pages}
  {241104} (\bibinfo {year} {2016})}\BibitemShut {NoStop}%
\bibitem [{\citenamefont {Chatziioannou}\ \emph {et~al.}(2012)\citenamefont
  {Chatziioannou}, \citenamefont {Yunes},\ and\ \citenamefont
  {Cornish}}]{PhysRevD.86.022004}%
  \BibitemOpen
  \bibfield  {author} {\bibinfo {author} {\bibfnamefont {Katerina}\
  \bibnamefont {Chatziioannou}}, \bibinfo {author} {\bibfnamefont {Nicol\'as}\
  \bibnamefont {Yunes}}, \ and\ \bibinfo {author} {\bibfnamefont {Neil}\
  \bibnamefont {Cornish}},\ }\bibfield  {title} {\enquote {\bibinfo {title}
  {Model-independent test of general relativity: An extended post-einsteinian
  framework with complete polarization content},}\ }\href {\doibase
  10.1103/PhysRevD.86.022004} {\bibfield  {journal} {\bibinfo  {journal} {Phys.
  Rev. D}\ }\textbf {\bibinfo {volume} {86}},\ \bibinfo {pages} {022004}
  (\bibinfo {year} {2012})}\BibitemShut {NoStop}%
\bibitem [{\citenamefont {Romero-Shaw}\ \emph
  {et~al.}(2023{\natexlab{b}})\citenamefont {Romero-Shaw}, \citenamefont
  {Gerosa},\ and\ \citenamefont {Loutrel}}]{Romero-Shaw:2022fbf}%
  \BibitemOpen
  \bibfield  {author} {\bibinfo {author} {\bibfnamefont {Isobel~M.}\
  \bibnamefont {Romero-Shaw}}, \bibinfo {author} {\bibfnamefont {Davide}\
  \bibnamefont {Gerosa}}, \ and\ \bibinfo {author} {\bibfnamefont {Nicholas}\
  \bibnamefont {Loutrel}},\ }\bibfield  {title} {\enquote {\bibinfo {title}
  {{Eccentricity or spin precession? Distinguishing subdominant effects in
  gravitational-wave data}},}\ }\href {\doibase 10.1093/mnras/stad031}
  {\bibfield  {journal} {\bibinfo  {journal} {Mon. Not. Roy. Astron. Soc.}\
  }\textbf {\bibinfo {volume} {519}},\ \bibinfo {pages} {5352--5357} (\bibinfo
  {year} {2023}{\natexlab{b}})},\ \Eprint {http://arxiv.org/abs/2211.07528}
  {arXiv:2211.07528 [astro-ph.HE]} \BibitemShut {NoStop}%
\bibitem [{\citenamefont {Ghosh}\ \emph {et~al.}(2016)\citenamefont {Ghosh}
  \emph {et~al.}}]{Ghosh:2016qgn}%
  \BibitemOpen
  \bibfield  {author} {\bibinfo {author} {\bibfnamefont {Abhirup}\ \bibnamefont
  {Ghosh}} \emph {et~al.},\ }\bibfield  {title} {\enquote {\bibinfo {title}
  {{Testing general relativity using golden black-hole binaries}},}\ }\href
  {\doibase 10.1103/PhysRevD.94.021101} {\bibfield  {journal} {\bibinfo
  {journal} {Phys. Rev. D}\ }\textbf {\bibinfo {volume} {94}},\ \bibinfo
  {pages} {021101} (\bibinfo {year} {2016})},\ \Eprint
  {http://arxiv.org/abs/1602.02453} {arXiv:1602.02453 [gr-qc]} \BibitemShut
  {NoStop}%
\bibitem [{\citenamefont {Abbott}\ \emph
  {et~al.}(2021{\natexlab{d}})\citenamefont {Abbott} \emph
  {et~al.}}]{LIGOScientific:2021sio}%
  \BibitemOpen
  \bibfield  {author} {\bibinfo {author} {\bibfnamefont {R.}~\bibnamefont
  {Abbott}} \emph {et~al.} (\bibinfo {collaboration} {LIGO Scientific, VIRGO,
  KAGRA}),\ }\bibfield  {title} {\enquote {\bibinfo {title} {{Tests of General
  Relativity with GWTC-3}},}\ }\href@noop {} {\  (\bibinfo {year}
  {2021}{\natexlab{d}})},\ \Eprint {http://arxiv.org/abs/2112.06861}
  {arXiv:2112.06861 [gr-qc]} \BibitemShut {NoStop}%
\bibitem [{\citenamefont {Romero-Shaw}\ \emph {et~al.}(2022)\citenamefont
  {Romero-Shaw}, \citenamefont {Lasky},\ and\ \citenamefont
  {Thrane}}]{RomeroShaw2022_multi_ecc}%
  \BibitemOpen
  \bibfield  {author} {\bibinfo {author} {\bibfnamefont {Isobel~M.}\
  \bibnamefont {Romero-Shaw}}, \bibinfo {author} {\bibfnamefont {Paul~D.}\
  \bibnamefont {Lasky}}, \ and\ \bibinfo {author} {\bibfnamefont {Eric}\
  \bibnamefont {Thrane}},\ }\bibfield  {title} {\enquote {\bibinfo {title}
  {{Four eccentric mergers increase the evidence that LIGO–Virgo–KAGRA’s
  binary black holes form dynamically}},}\ }\href {\doibase
  10.3847/1538-4357/ac9798} {\bibfield  {journal} {\bibinfo  {journal} {The
  Astrophysical Journal}\ }\textbf {\bibinfo {volume} {940}},\ \bibinfo {pages}
  {119} (\bibinfo {year} {2022})},\ \Eprint {http://arxiv.org/abs/2206.14695}
  {arXiv:2206.14695 [astro-ph.HE]} \BibitemShut {NoStop}%
\bibitem [{\citenamefont {Romero-Shaw}\ \emph {et~al.}(2020)\citenamefont
  {Romero-Shaw}, \citenamefont {Lasky}, \citenamefont {Thrane},\ and\
  \citenamefont {Bustillo}}]{RomeroShaw2020_GW190521_eccentric}%
  \BibitemOpen
  \bibfield  {author} {\bibinfo {author} {\bibfnamefont {Isobel~M.}\
  \bibnamefont {Romero-Shaw}}, \bibinfo {author} {\bibfnamefont {Paul~D.}\
  \bibnamefont {Lasky}}, \bibinfo {author} {\bibfnamefont {Eric}\ \bibnamefont
  {Thrane}}, \ and\ \bibinfo {author} {\bibfnamefont {Juan~Calder\'on}\
  \bibnamefont {Bustillo}},\ }\bibfield  {title} {\enquote {\bibinfo {title}
  {{GW190521: orbital eccentricity and signatures of dynamical formation in a
  binary black hole merger signal}},}\ }\href {\doibase
  10.3847/2041-8213/abbe26} {\bibfield  {journal} {\bibinfo  {journal} {The
  Astrophysical Journal Letters}\ }\textbf {\bibinfo {volume} {903}},\ \bibinfo
  {pages} {L5} (\bibinfo {year} {2020})}\BibitemShut {NoStop}%
\bibitem [{\citenamefont {Romero-Shaw}\ \emph {et~al.}(2025)\citenamefont
  {Romero-Shaw}, \citenamefont {Stegmann}, \citenamefont {Tagawa},
  \citenamefont {Gerosa}, \citenamefont {Samsing}, \citenamefont {Gupte},\ and\
  \citenamefont {Green}}]{Romero-Shaw:2025vbc}%
  \BibitemOpen
  \bibfield  {author} {\bibinfo {author} {\bibfnamefont {Isobel}\ \bibnamefont
  {Romero-Shaw}}, \bibinfo {author} {\bibfnamefont {Jakob}\ \bibnamefont
  {Stegmann}}, \bibinfo {author} {\bibfnamefont {Hiromichi}\ \bibnamefont
  {Tagawa}}, \bibinfo {author} {\bibfnamefont {Davide}\ \bibnamefont {Gerosa}},
  \bibinfo {author} {\bibfnamefont {Johan}\ \bibnamefont {Samsing}}, \bibinfo
  {author} {\bibfnamefont {Nihar}\ \bibnamefont {Gupte}}, \ and\ \bibinfo
  {author} {\bibfnamefont {Stephen~R.}\ \bibnamefont {Green}},\ }\bibfield
  {title} {\enquote {\bibinfo {title} {{GW200208{\_}222617 as an eccentric
  black-hole binary merger: properties and astrophysical implications}},}\
  }\href@noop {} {\  (\bibinfo {year} {2025})},\ \Eprint
  {http://arxiv.org/abs/2506.17105} {arXiv:2506.17105 [astro-ph.HE]}
  \BibitemShut {NoStop}%
\bibitem [{\citenamefont {de~Lluc~Planas}\ \emph {et~al.}(2025)\citenamefont
  {de~Lluc~Planas}, \citenamefont {Husa}, \citenamefont {Ramos-Buades},\ and\
  \citenamefont {Valencia}}]{Planas2025_GW200105_IMR_ecc}%
  \BibitemOpen
  \bibfield  {author} {\bibinfo {author} {\bibfnamefont {Maria}\ \bibnamefont
  {de~Lluc~Planas}}, \bibinfo {author} {\bibfnamefont {Sascha}\ \bibnamefont
  {Husa}}, \bibinfo {author} {\bibfnamefont {Antoni}\ \bibnamefont
  {Ramos-Buades}}, \ and\ \bibinfo {author} {\bibfnamefont {Jorge}\
  \bibnamefont {Valencia}},\ }\bibfield  {title} {\enquote {\bibinfo {title}
  {{First eccentric inspiral-merger-ringdown analysis of neutron star-black
  hole mergers}},}\ }\href {\doibase 10.48550/arXiv.2506.01760} {\bibfield
  {journal} {\bibinfo  {journal} {arXiv preprint arXiv:2506.01760}\ } (\bibinfo
  {year} {2025}),\ 10.48550/arXiv.2506.01760}\BibitemShut {NoStop}%
\bibitem [{\citenamefont {Gupte}\ \emph {et~al.}(2024)\citenamefont {Gupte},
  \citenamefont {Ramos-Buades}, \citenamefont {Buonanno}, \citenamefont {Gair},
  \citenamefont {Miller}, \citenamefont {Dax}, \citenamefont {Green},
  \citenamefont {Pürrer}, \citenamefont {Wildberger}, \citenamefont {Macke},
  \citenamefont {Romero-Shaw},\ and\ \citenamefont
  {Schölkopf}}]{Gupte2024_pop_ecc}%
  \BibitemOpen
  \bibfield  {author} {\bibinfo {author} {\bibfnamefont {Nihar}\ \bibnamefont
  {Gupte}}, \bibinfo {author} {\bibfnamefont {Antoni}\ \bibnamefont
  {Ramos-Buades}}, \bibinfo {author} {\bibfnamefont {Alessandra}\ \bibnamefont
  {Buonanno}}, \bibinfo {author} {\bibfnamefont {Jonathan}\ \bibnamefont
  {Gair}}, \bibinfo {author} {\bibfnamefont {M.~Coleman}\ \bibnamefont
  {Miller}}, \bibinfo {author} {\bibfnamefont {Maximilian}\ \bibnamefont
  {Dax}}, \bibinfo {author} {\bibfnamefont {Stephen~R.}\ \bibnamefont {Green}},
  \bibinfo {author} {\bibfnamefont {Michael}\ \bibnamefont {Pürrer}}, \bibinfo
  {author} {\bibfnamefont {Jonas}\ \bibnamefont {Wildberger}}, \bibinfo
  {author} {\bibfnamefont {Jakob}\ \bibnamefont {Macke}}, \bibinfo {author}
  {\bibfnamefont {Isobel~M.}\ \bibnamefont {Romero-Shaw}}, \ and\ \bibinfo
  {author} {\bibfnamefont {Bernhard}\ \bibnamefont {Schölkopf}},\ }\bibfield
  {title} {\enquote {\bibinfo {title} {{Evidence for eccentricity in the
  population of binary black holes observed by LIGO–Virgo–KAGRA}},}\ }\href
  {\doibase 10.48550/arXiv.2404.14286} {\bibfield  {journal} {\bibinfo
  {journal} {arXiv preprint arXiv:2404.14286}\ } (\bibinfo {year} {2024}),\
  10.48550/arXiv.2404.14286}\BibitemShut {NoStop}%
\bibitem [{\citenamefont {Dhurkunde}\ and\ \citenamefont
  {et~al.}(2025)}]{DhurkundeNitz2025_eccSearch_O3}%
  \BibitemOpen
  \bibfield  {author} {\bibinfo {author} {\bibfnamefont {Rahul}\ \bibnamefont
  {Dhurkunde}}\ and\ \bibinfo {author} {\bibfnamefont {Alexander H.~Nitz}\
  \bibnamefont {et~al.}},\ }\bibfield  {title} {\enquote {\bibinfo {title}
  {{Search for eccentric NSBH and BNS mergers in the third observing run of
  Advanced LIGO and Virgo}},}\ }\href {\doibase 10.1103/PhysRevD.111.103018}
  {\bibfield  {journal} {\bibinfo  {journal} {Physical Review D}\ }\textbf
  {\bibinfo {volume} {111}},\ \bibinfo {pages} {103018} (\bibinfo {year}
  {2025})}\BibitemShut {NoStop}%
\bibitem [{\citenamefont {Kacanja}\ \emph {et~al.}(2025)\citenamefont
  {Kacanja}, \citenamefont {Soni},\ and\ \citenamefont
  {Nitz}}]{Kacanja:2025kpr}%
  \BibitemOpen
  \bibfield  {author} {\bibinfo {author} {\bibfnamefont {Keisi}\ \bibnamefont
  {Kacanja}}, \bibinfo {author} {\bibfnamefont {Kanchan}\ \bibnamefont {Soni}},
  \ and\ \bibinfo {author} {\bibfnamefont {Alexander~Harvey}\ \bibnamefont
  {Nitz}},\ }\bibfield  {title} {\enquote {\bibinfo {title} {{Eccentricity
  signatures in LIGO-Virgo-KAGRA's BNS and NSBH binaries}},}\ }\href@noop {} {\
   (\bibinfo {year} {2025})},\ \Eprint {http://arxiv.org/abs/2508.00179}
  {arXiv:2508.00179 [gr-qc]} \BibitemShut {NoStop}%
\bibitem [{\citenamefont {Gayathri}\ \emph {et~al.}(2020)\citenamefont
  {Gayathri}, \citenamefont {Healy}, \citenamefont {Lange}, \citenamefont
  {O'Brien}, \citenamefont {Szczepanczyk}, \citenamefont {Bartos},
  \citenamefont {Campanelli}, \citenamefont {Klimenko}, \citenamefont
  {Lousto},\ and\ \citenamefont {O'Shaughnessy}}]{Gayathri2020_GW190521_NR}%
  \BibitemOpen
  \bibfield  {author} {\bibinfo {author} {\bibfnamefont {V.}~\bibnamefont
  {Gayathri}}, \bibinfo {author} {\bibfnamefont {J.}~\bibnamefont {Healy}},
  \bibinfo {author} {\bibfnamefont {J.}~\bibnamefont {Lange}}, \bibinfo
  {author} {\bibfnamefont {B.}~\bibnamefont {O'Brien}}, \bibinfo {author}
  {\bibfnamefont {M.}~\bibnamefont {Szczepanczyk}}, \bibinfo {author}
  {\bibfnamefont {I.}~\bibnamefont {Bartos}}, \bibinfo {author} {\bibfnamefont
  {M.}~\bibnamefont {Campanelli}}, \bibinfo {author} {\bibfnamefont
  {S.}~\bibnamefont {Klimenko}}, \bibinfo {author} {\bibfnamefont
  {C.}~\bibnamefont {Lousto}}, \ and\ \bibinfo {author} {\bibfnamefont
  {R.}~\bibnamefont {O'Shaughnessy}},\ }\bibfield  {title} {\enquote {\bibinfo
  {title} {{Eccentricity estimate for black hole mergers with numerical
  relativity simulations}},}\ }\href {\doibase 10.48550/arXiv.2009.05461}
  {\bibfield  {journal} {\bibinfo  {journal} {arXiv preprint arXiv:2009.05461}\
  } (\bibinfo {year} {2020}),\ 10.48550/arXiv.2009.05461}\BibitemShut {NoStop}%
\bibitem [{\citenamefont {Miller}\ \emph {et~al.}(2023)\citenamefont {Miller},
  \citenamefont {Isi}, \citenamefont {Chatziioannou}, \citenamefont {Varma},\
  and\ \citenamefont {Mandel}}]{Miller2023_GW190521_precessing}%
  \BibitemOpen
  \bibfield  {author} {\bibinfo {author} {\bibfnamefont {Simona~J.}\
  \bibnamefont {Miller}}, \bibinfo {author} {\bibfnamefont {Maximiliano}\
  \bibnamefont {Isi}}, \bibinfo {author} {\bibfnamefont {Katerina}\
  \bibnamefont {Chatziioannou}}, \bibinfo {author} {\bibfnamefont {Vijay}\
  \bibnamefont {Varma}}, \ and\ \bibinfo {author} {\bibfnamefont {Ilya}\
  \bibnamefont {Mandel}},\ }\bibfield  {title} {\enquote {\bibinfo {title}
  {{GW190521: tracing imprints of spin‑precession on the most massive black
  hole binary}},}\ }\href {\doibase 10.48550/arXiv.2310.01544} {\bibfield
  {journal} {\bibinfo  {journal} {arXiv preprint arXiv:2310.01544}\ } (\bibinfo
  {year} {2023}),\ 10.48550/arXiv.2310.01544}\BibitemShut {NoStop}%
\bibitem [{\citenamefont {Bustillo}\ \emph
  {et~al.}(2020{\natexlab{a}})\citenamefont {Bustillo}, \citenamefont
  {Sanchis‑Gual}, \citenamefont {Torres‑Forn{\'e}},\ and\ \citenamefont
  {Font}}]{CalderonBustillo2020_headOnGW190521}%
  \BibitemOpen
  \bibfield  {author} {\bibinfo {author} {\bibfnamefont {Juan~Calder{\'o}n}\
  \bibnamefont {Bustillo}}, \bibinfo {author} {\bibfnamefont {Nicolas}\
  \bibnamefont {Sanchis‑Gual}}, \bibinfo {author} {\bibfnamefont {Alejandro}\
  \bibnamefont {Torres‑Forn{\'e}}}, \ and\ \bibinfo {author} {\bibfnamefont
  {Jos{\'e}~A.}\ \bibnamefont {Font}},\ }\bibfield  {title} {\enquote {\bibinfo
  {title} {{Confusing head‑on collisions with precessing intermediate‑mass
  binary black hole mergers}},}\ }\href {\doibase 10.48550/arXiv.2009.01066}
  {\bibfield  {journal} {\bibinfo  {journal} {arXiv preprint arXiv:2009.01066}\
  } (\bibinfo {year} {2020}{\natexlab{a}}),\
  10.48550/arXiv.2009.01066}\BibitemShut {NoStop}%
\bibitem [{\citenamefont {Bustillo}\ \emph
  {et~al.}(2020{\natexlab{b}})\citenamefont {Bustillo}, \citenamefont
  {Sanchis‑Gual}, \citenamefont {Torres‑Forn{\'e}}, \citenamefont {Font},
  \citenamefont {Vajpeyi}, \citenamefont {Smith}, \citenamefont {Herdeiro},
  \citenamefont {Radu},\ and\ \citenamefont
  {Leong}}]{CalderonBustillo2020_ProcaHeadOn}%
  \BibitemOpen
  \bibfield  {author} {\bibinfo {author} {\bibfnamefont {Juan~Calder{\'o}n}\
  \bibnamefont {Bustillo}}, \bibinfo {author} {\bibfnamefont {Nicolas}\
  \bibnamefont {Sanchis‑Gual}}, \bibinfo {author} {\bibfnamefont {Alejandro}\
  \bibnamefont {Torres‑Forn{\'e}}}, \bibinfo {author} {\bibfnamefont
  {Jos{\'e}~A.}\ \bibnamefont {Font}}, \bibinfo {author} {\bibfnamefont {Avi}\
  \bibnamefont {Vajpeyi}}, \bibinfo {author} {\bibfnamefont {Rory}\
  \bibnamefont {Smith}}, \bibinfo {author} {\bibfnamefont {Carlos}\
  \bibnamefont {Herdeiro}}, \bibinfo {author} {\bibfnamefont {Eugen}\
  \bibnamefont {Radu}}, \ and\ \bibinfo {author} {\bibfnamefont {Samson H.~W.}\
  \bibnamefont {Leong}},\ }\bibfield  {title} {\enquote {\bibinfo {title}
  {{GW190521 as a merger of Proca stars: a potential new vector boson of
  $8.7\times10^{-13}\,$eV}},}\ }\href {\doibase 10.48550/arXiv.2009.05376}
  {\bibfield  {journal} {\bibinfo  {journal} {arXiv preprint arXiv:2009.05376}\
  } (\bibinfo {year} {2020}{\natexlab{b}}),\
  10.48550/arXiv.2009.05376}\BibitemShut {NoStop}%
\bibitem [{\citenamefont {{Mishra}}\ \emph {et~al.}(2021)\citenamefont
  {{Mishra}}, \citenamefont {{Meena}}, \citenamefont {{More}}, \citenamefont
  {{Bose}},\ and\ \citenamefont {{Bagla}}}]{2021MNRAS.508.4869M}%
  \BibitemOpen
  \bibfield  {author} {\bibinfo {author} {\bibfnamefont {Anuj}\ \bibnamefont
  {{Mishra}}}, \bibinfo {author} {\bibfnamefont {Ashish~Kumar}\ \bibnamefont
  {{Meena}}}, \bibinfo {author} {\bibfnamefont {Anupreeta}\ \bibnamefont
  {{More}}}, \bibinfo {author} {\bibfnamefont {Sukanta}\ \bibnamefont
  {{Bose}}}, \ and\ \bibinfo {author} {\bibfnamefont {Jasjeet~Singh}\
  \bibnamefont {{Bagla}}},\ }\bibfield  {title} {\enquote {\bibinfo {title}
  {{Gravitational lensing of gravitational waves: effect of microlens
  population in lensing galaxies}},}\ }\href {\doibase 10.1093/mnras/stab2875}
  {\bibfield  {journal} {\bibinfo  {journal} {Monthly Notices of the Royal
  Astronomical Society}\ }\textbf {\bibinfo {volume} {508}},\ \bibinfo {pages}
  {4869--4886} (\bibinfo {year} {2021})},\ \Eprint
  {http://arxiv.org/abs/2102.03946} {arXiv:2102.03946 [astro-ph.CO]}
  \BibitemShut {NoStop}%
\bibitem [{\citenamefont {Reitze}\ \emph {et~al.}(2019)\citenamefont {Reitze}
  \emph {et~al.}}]{Reitze:2019iox}%
  \BibitemOpen
  \bibfield  {author} {\bibinfo {author} {\bibfnamefont {David}\ \bibnamefont
  {Reitze}} \emph {et~al.},\ }\bibfield  {title} {\enquote {\bibinfo {title}
  {{Cosmic Explorer: The U.S. Contribution to Gravitational-Wave Astronomy
  beyond LIGO}},}\ }\href@noop {} {\bibfield  {journal} {\bibinfo  {journal}
  {Bull. Am. Astron. Soc.}\ }\textbf {\bibinfo {volume} {51}},\ \bibinfo
  {pages} {035} (\bibinfo {year} {2019})},\ \Eprint
  {http://arxiv.org/abs/1907.04833} {arXiv:1907.04833 [astro-ph.IM]}
  \BibitemShut {NoStop}%
\bibitem [{\citenamefont {Punturo}\ \emph {et~al.}(2010)\citenamefont
  {Punturo}, \citenamefont {Abernathy}, \citenamefont {Acernese}, \citenamefont
  {Allen}, \citenamefont {Andersson}, \citenamefont {Arun}, \citenamefont
  {Barone}, \citenamefont {Barr}, \citenamefont {Barsuglia}, \citenamefont
  {Beker} \emph {et~al.}}]{punturo2010}%
  \BibitemOpen
  \bibfield  {author} {\bibinfo {author} {\bibfnamefont {M}~\bibnamefont
  {Punturo}}, \bibinfo {author} {\bibfnamefont {M}~\bibnamefont {Abernathy}},
  \bibinfo {author} {\bibfnamefont {F}~\bibnamefont {Acernese}}, \bibinfo
  {author} {\bibfnamefont {B}~\bibnamefont {Allen}}, \bibinfo {author}
  {\bibfnamefont {Nils}\ \bibnamefont {Andersson}}, \bibinfo {author}
  {\bibfnamefont {K}~\bibnamefont {Arun}}, \bibinfo {author} {\bibfnamefont
  {F}~\bibnamefont {Barone}}, \bibinfo {author} {\bibfnamefont {B}~\bibnamefont
  {Barr}}, \bibinfo {author} {\bibfnamefont {M}~\bibnamefont {Barsuglia}},
  \bibinfo {author} {\bibfnamefont {M}~\bibnamefont {Beker}},  \emph {et~al.},\
  }\bibfield  {title} {\enquote {\bibinfo {title} {The einstein telescope: a
  third-generation gravitational wave observatory},}\ }\href@noop {} {\bibfield
   {journal} {\bibinfo  {journal} {Classical and Quantum Gravity}\ }\textbf
  {\bibinfo {volume} {27}},\ \bibinfo {pages} {194002} (\bibinfo {year}
  {2010})}\BibitemShut {NoStop}%
\bibitem [{\citenamefont {Amaro-Seoane}\ \emph {et~al.}(2017)\citenamefont
  {Amaro-Seoane} \emph {et~al.}}]{LISA:2017pwj}%
  \BibitemOpen
  \bibfield  {author} {\bibinfo {author} {\bibfnamefont {Pau}\ \bibnamefont
  {Amaro-Seoane}} \emph {et~al.} (\bibinfo {collaboration} {LISA}),\ }\bibfield
   {title} {\enquote {\bibinfo {title} {{Laser Interferometer Space
  Antenna}},}\ }\href@noop {} {\  (\bibinfo {year} {2017})},\ \Eprint
  {http://arxiv.org/abs/1702.00786} {arXiv:1702.00786 [astro-ph.IM]}
  \BibitemShut {NoStop}%
\bibitem [{\citenamefont {Sato}\ \emph {et~al.}(2017)\citenamefont {Sato} \emph
  {et~al.}}]{Sato:2017dkf}%
  \BibitemOpen
  \bibfield  {author} {\bibinfo {author} {\bibfnamefont {Shuichi}\ \bibnamefont
  {Sato}} \emph {et~al.},\ }\bibfield  {title} {\enquote {\bibinfo {title}
  {{The status of DECIGO}},}\ }\href {\doibase 10.1088/1742-6596/840/1/012010}
  {\bibfield  {journal} {\bibinfo  {journal} {J. Phys. Conf. Ser.}\ }\textbf
  {\bibinfo {volume} {840}},\ \bibinfo {pages} {012010} (\bibinfo {year}
  {2017})}\BibitemShut {NoStop}%
\end{thebibliography}%
\end{document}